\documentclass{pasa}%

\usepackage{graphicx}
\usepackage[para]{threeparttable}
\usepackage{subfigure}
\usepackage{wrapfig,lipsum}
\usepackage{rotating}
\usepackage{amsmath}
\usepackage{siunitx}
\usepackage{upgreek}
\usepackage{diagbox}
\usepackage{tabularx}
\newcommand\setrow[1]{\gdef\rowmac{#1}#1\ignorespaces}
\newcommand\clearrow{\global\let\rowmac\relax}
\clearrow
\usepackage{slashbox,multirow}

\title[MWA GW searches]{MWA rapid follow-up of gravitational wave transients: prospects for detecting prompt radio counterparts}

\author[J. Tian et al.]{J. Tian$^{1}$\thanks{E-mail: jun.tian@postgrad.curtin.edu.au}, G. E. Anderson$^1$, A. J. Cooper$^{11,2,3}$, K. Gourdji$^{4,5}$, M. Sokolowski$^1$, A. Rowlinson$^{2,3}$, A. Williams$^1$, G. Sleap$^1$, D. Dobie$^{4,5}$, D. L. Kaplan$^6$, Tara Murphy$^{5,7}$, S. J. Tingay$^1$, F. H. Panther$^{5,8}$, P. D. Lasky$^{5,9}$, A. Bahramian$^1$, J. C. A. Miller-Jones$^1$, C. W. James$^1$, B. W. Meyers$^1$, S. J. McSweeney$^1$, P. J. Hancock$^{10}$.
\affil{$^1$International Centre for Radio Astronomy Research, Curtin University, GPO Box U1987, Perth, WA 6845, Australia}%
\affil{$^2$Anton Pannekoek Institute, University of Amsterdam, Postbus 94249, 1090 GE, Amsterdam, The Netherlands}
\affil{$^3$ASTRON, the Netherlands Institute for Radio Astronomy, Oude Hoogeveensedijk 4, 7991 PD, Dwingeloo, The Netherlands}
\affil{$^4$Centre for Astrophysics and Supercomputing, Swinburne University of Technology, Hawthorn VIC 3122, Australia}
\affil{$^5$OzGrav: ARC Centre of Excellence for Gravitational Wave Discovery, Hawthorn VIC 3122, Australia}
\affil{$^6$Department of Physics, University of Wisconsin-Milwaukee, 1900 E. Kenwood Boulevard, Milwaukee, WI 53211, USA}
\affil{$^7$Sydney Institute for Astronomy, School of Physics, The University of Sydney, NSW 2006, Australia}
\affil{$^8$Department of Physics, University of Western Australia, Crawley WA 6009, Australia}
\affil{$^9$School of Physics and Astronomy, Monash University, VIC 3800}
\affil{$^{10}$Curtin Institute for Computation, Curtin University, GPO Box U1987, Perth, 6845, WA, Australia}
\affil{$^{11}$Astrophysics, The University of Oxford, Keble Road, Oxford OX1 3RH, UK}
}%

\jid{PASA}
\doi{10.1017/pas.\the\year.xxx}
\jyear{\the\year}

\usepackage{aas_macros}
\usepackage{hyperref} 
\hypersetup{colorlinks,citecolor=blue,linkcolor=blue,urlcolor=blue}

\begin{document}

\begin{frontmatter}
\maketitle

\begin{abstract}
We present and evaluate the prospects for detecting coherent radio counterparts to gravitational wave (GW) events using Murchison Widefield Array (MWA) triggered observations. The MWA rapid-response system, combined with its buffering mode ($\sim4$ minutes negative latency), enables us to catch any radio signals produced from seconds prior to hours after a binary neutron star (BNS) merger. The large field of view of the MWA ($\sim1000\,\text{deg}^2$ at 120\,MHz) and its location under the high sensitivity sky region of the LIGO-Virgo-KAGRA (LVK) detector network, forecast a high chance of being on-target for a GW event. We consider three observing configurations for the MWA to follow up GW BNS merger events, including a single dipole per tile, the full array, and four sub-arrays. We then perform a population synthesis of BNS systems to predict the radio detectable fraction of GW events using these configurations. We find that the configuration with four sub-arrays is the best compromise between sky coverage and sensitivity as it is capable of placing meaningful constraints on the radio emission from 12.6\% of GW BNS detections. Based on the timescales of four BNS merger coherent radio emission models, we propose an observing strategy that involves triggering the buffering mode to target coherent signals emitted prior to, during or shortly following the merger, which is then followed by continued recording for up to three hours to target later time post-merger emission. We expect MWA to trigger on $\sim5\text{--}22$ BNS merger events during the LVK O4 observing run, which could potentially result in two detections of predicted coherent emission.
\end{abstract}

\end{frontmatter}

\section{Introduction}

\begin{figure*}
\centering
\includegraphics[width=0.7\textwidth]{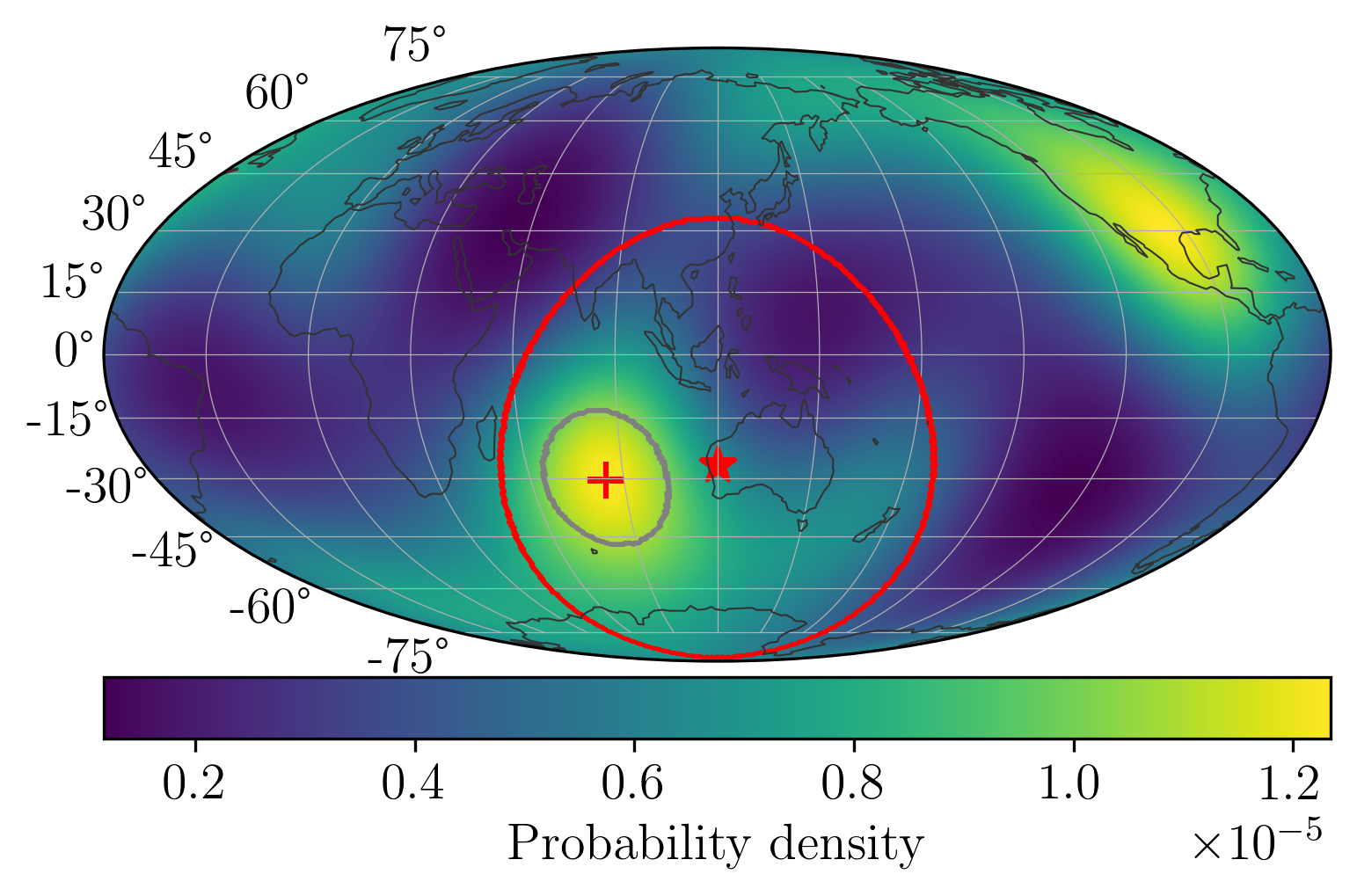}
\caption{The LVK GW sensitivity map for O4 projected on the Earth.
We used the sensitivity map generated by the LALSuite software suite \citep{lalsuite}, and assumed the same distribution of signal-to-noise ratio (S/N) of GW signals as simulated for O3 \citep{LIGO21b, LIGO21}. 
The colour scale corresponds to the probability of detecting a GW event at a particular sky position with respect to the Earth. 
The highest sensitivity region in the Southern Hemisphere (marked by a red plus) is at an elevation of $\mathbf{58.5^\circ}$ in the MWA field (also discussed in \citealt{Wang20}).
The red star marks the location of MWA, the red contour shows the full sky coverage of MWA down to an elevation of $\mathbf{30^\circ}$, and the grey contour shows the FoV of a standard MWA pointing centred on the highest sensitivity region down to 20\% of the primary beam at 120\,MHz.
This map demonstrates that MWA is well placed to observe the highest sensitivity region of GW detection in the Southern Hemisphere,
with 30.5\% and 4.9\% of events expected to be within the red and grey contours, respectively (see Section~\ref{sec:GW_map}).}
\label{GW_map}
\end{figure*}

On 2015 September 14, the LIGO–Virgo–KAGRA Collaboration (LVK) detected the first gravitational wave (GW) signal, GW150914, from a binary black hole (BBH) merger, marking the start of a new era in astronomy \citep{Abbott16a}. Since then, more GW signals have been detected, with most originating from BBH mergers \citep{Abbott16b, Abbott19}, two from binary neutron star (BNS) mergers \citep{Abbott17, Abbott20b}, and two to four from BH-NS mergers \citep{Abbott21a, Abbott21b} thanks to the commissioning of the Advanced LIGO and Virgo Interferometer (aLIGO/Virgo; \citealt{Aasi15, Acernese15}). 
Remarkably, electromagnetic (EM) counterparts of GWs were identified for a BNS merger \citep{Abbott17b}.

The contemporaneous detection of GW170817 and short gamma-ray burst (GRB) 170817A \citep{Abbott17, Abbott17b, Goldstein17} significantly increased the utility of GW signals and ignited a campaign of multi-wavelength follow-up. This led to observations in almost every EM band, yielding a wealth of information on compact binary merger physics including short GRB mechanisms \citep{Mooley18b, Nakar18} and the NS equation of state (e.g., \citealt{Abbott18, Raithel18}). However, GW170817 is the only gravitational wave-detected event with confirmed joint EM detections to date, although there was substantial effort devoted to following up GWs (e.g. \citealt{Coughlin19, Graham20, Alexander21, Dobie22, Panther23}). While identifying EM counterparts is extremely useful for studying GW physics, a few factors such as the delay in issuing GW alerts and the error region of hundreds to thousands of square degrees (especially for early warning alerts) mean it is a challenging task (e.g. \citealt{Kasliwal14, Cowperthwaite15}).

Among the EM counterparts associated with GW transients is the theorised coherent radio emission (e.g. \citealt{Platts19, Rowlinson19, Cooper22}). Many models predict prompt, fast radio burst (FRB) like signals or persistent pulsar-like emission in the course of compact binary mergers. While BH-NS mergers could produce some coherent radio emission, we are focusing on BNS mergers in this paper. The earliest radio emission could come from the inspiral phase, where interactions of the NS magnetic fields just preceding the merger could revive the pulsar emission mechanism \citep{Lyutikov19}. The merger may launch an extremely relativistic jet, interacting with the interstellar medium (ISM), that produces an FRB-like signal \citep{Usov00}. If the merger remnant is a supramassive \textbf{(i.e. mass larger than the maximum mass allowed for a static NS)}, rapidly rotating, highly magnetised NS (from hereon referred to as a magnetar), we may expect pulsar-like emission powered by dipole magnetic braking during the lifetime of
the magnetar \citep{Totani13} and/or magnetically powered radio bursts from the magnetar remnant \citep{Lyubarsky14}. Finally, as the magnetar remnant spins down, it may collapse into a BH ejecting its magnetosphere and producing a prompt radio burst \citep{Falcke14, Zhang14}.

There have been several searches for prompt coherent radio counterparts to GW transients \citep{Andreoni17, Callister19, Artkop19, Bhakta21, LVK22, Moroianu23}. 
The Murchison Widefield Array (MWA; \citealt{Tingay13, Wayth18}) participated in the Australian-led multi-wavelength follow-up program of GW sources, to search for coherent radio emission associated with GW170817 within five days of the GW trigger, but no signals were observed above 51\,mJy on 150\,min timescales \citep{Andreoni17}.
\citet{Callister19} used the Owens Valley Radio Observatory Long Wavelength Array (OVRO-LWA) to search a $\sim900\,\text{deg}^2$ region for prompt radio transients between 27--84\,MHz within the positional error of a BBH merger GW170104 ($\sim1600\,\text{deg}^2$) six hours after the GW detection, and obtained a typical upper limit of 2.4\,Jy on 13\,s timescales. Similar searches at higher frequencies were conducted with better sensitivity but in a much smaller search area. Using the Karl G. Jansky Very Large Array (VLA), \citet{Artkop19} and \citet{Bhakta21} searched only small regions ($<1\,\text{deg}^2$) of possible gamma-ray counterparts identified in the GW localisation area days after the GW arrival (also from BBH mergers), resulting in an upper limit of $450\,\mu$Jy on 1\,hr timescales at 1.4\,GHz and $75\,\mu$Jy on 3\,hr timescales at 6\,GHz, respectively. 
Very recently, \citet{Moroianu23} conducted a search for GW-FRB associations by cross matching the first CHIME/FRB catalogue \citep{CHIME21} with the GW sources detected in the first half of the third GW observing run (O3a; \citealt{Abbott21a}), and reported a potential association, i.e. FRB 20190425A occurred 2.5\,hr following GW190425 and within the GW sky localisation area, though at a low significance of $2.8\sigma$.

With the LVK O4 observing run \citep{Abbott18b} commencing, we are now presented with an unprecedented opportunity to search for the theorised coherent radio emission associated with BNS mergers.
The lack of strong associations between BNS mergers and coherent radio emission in previous studies may be due to several factors, including the radio telescope having an insufficient field of view for covering the large uncertainty regions of GW events, a large delay between the GW detection and the radio follow-up, 
or insufficient sensitivity. 
In order to combat these issues, 
we present an observing strategy for searching for coherent radio counterparts to GW transients with the MWA.

The MWA operates over a frequency range between 80 and 300\,MHz, with an instantaneous bandwidth of 30.72\,MHz, and a field of view (FoV) ranging from $\sim300-1000$\,deg$^2$ \citep{Tingay13}. It is suitable for finding prompt radio counterparts to GWs thanks to a few features. First, we have a unique opportunity as the MWA is well placed to target the highest sensitivity zone of the GW detector network over the Indian Ocean, as shown in Figure~\ref{GW_map}. 
Second is its large FoV. Given the poor localisation of GW events, especially for pre-merger detections ($\sim2000\,\text{deg}^2$ expected for O4)
\footnote{\url{https://emfollow.docs.ligo.org/userguide/}}, the MWA is able to cover a large proportion of the GW positional 
error regions. 
Third, the MWA has a rapid-response observing mode that can follow up a transient detection within 30\,s
of receiving an alert (\citealt{Kaplan15,hancock19,Anderson20, Tian22a, Tian22b}), and is now capable of storing high time resolution ($781.25$\,ns) data in a ring buffer that can be used to search for signals up to 240\,s prior to receiving an alert \citep{Morrison23}.
\textbf{For the utility of the ring buffer in the context of detecting coherent radio emission from BNS mergers see Section~\ref{sec:strategies}.}
This, combined with the dispersive delay expected at the MWA observing frequencies, allows us to capture the earliest radio signals predicted to be produced by BNS mergers. Furthermore, the MWA can trigger on transient alerts with the Voltage Capture System (VCS; \citealt{Morrison23}), which enables the capture of Nyquist-sampled voltage data. The desired time and frequency resolution can then be defined by the use case, i.e., some combination of frequency and time binning between 1.28\,MHz/781.25\,ns, and 1\,Hz/1\,s.

Given the above advantages, in this paper we discuss the prospect of detecting prompt radio emission from GW events with the MWA. Possible observing strategies for the MWA have already been investigated by \citet{Kaplan16} and \citet{James19}. However, these works do not consider specific emission models and their detectability. Here we focus on our success rate based on the model predictions applicable to BNS mergers in the context of the LVK O4 observing run \citep{Abbott18b}. 
We need to consider two problems in order to maximise our success of detecting prompt radio emission from BNS mergers: how the viewing angle of these mergers affects our chance of detection; 
and how the MWA can overcome a significant limitation for observatories with smaller FoV -- the ability to follow up the most poorly localised GW events, especially those that may be identified pre-merger from the gravitational waves emitted during the inspiral (e.g. \citealt{Sachdev20, Kovalam22}).
The goal of this paper is to devise the optimal observing strategy based on our investigation of these two problems.

In Section~\ref{sec:models}, we review some of the theoretical models that predict coherent radio emission to be produced by BNS mergers and how they are affected by inclination angle along our line of sight. In Section~\ref{sec:simulation}, we perform a population synthesis of GW sources and provide GW and radio detection criteria for deducing the jointly detectable population. We also calculate radio detection rates of GW events by taking into account the predicted distribution of GW detections in the sky and sensitivity variations of the MWA over different pointing directions. In Section~\ref{sec:strategies}, we propose a two-pronged triggering strategy for the MWA to follow up GW events based on the time frame that each of the coherent emission models are likely to occur during a BNS merger.

\section{Coherent emission from BNS mergers}\label{sec:models}

A number of models predict that BNS mergers could give rise to coherent radio emission (for a review see \citealt{Rowlinson19}), which could potentially be detected using MWA rapid-response observations
of GW transients.
In this Section, we revisit the fluence or flux density predictions of these emission models but also accounting for 
the BNS merger viewing angle (i.e. the angle between the observer's line of sight and the orbital angular momentum) up to a maximum distance of 190\,Mpc, which is the nominal horizon limit for O4 \citep{Abbott20}. 
Note that BH-NS mergers are not discussed here because several of the emission models are not relevant, including the interaction of NS magnetic fields (which is not possible with just one NS; see Section~\ref{sec:model_NSB}), and the magnetar collapse model (see Section~\ref{sec:model_col}) as we expect the BH-NS to directly collapse to a BH upon merger.

\subsection{Interactions of NS magnetic fields}\label{sec:model_NSB}

The earliest coherent radio emission produced by BNS mergers may occur during 
the inspiral phase, 
when the magnetospheric interaction between the two NSs could revive an enhanced pulsar emission mechanism \citep{Lipunov96,Metzger16}. In order to derive the luminosity of the pre-merger emission, here we consider a simple scenario that in the binary system one NS is highly magnetised (the primary NS) and the other moves in the magnetic field of the primary NS like a perfect conductor due to negligible magnetisation (the secondary NS; \citealt{Lyutikov19, Cooper22}). The pre-merger emission stems from an electric field induced by the motion of the secondary NS, which has a significant component parallel to the magnetic field, which accelerates particles. This parallel electric field $E_{\parallel}$ increases as the binary separation $a(t)$ shrinks due to gravitational radiation, and is given by \citep{Cooper22}

\begin{equation}
    E_{\parallel}=f(r, \theta, \phi)B(r, \theta, \phi)\beta,
\end{equation}

\noindent where $(r, \theta, \phi)$ is a spherical coordinate system centered on the secondary NS, $f(r, \theta, \phi)$ is a position dependent prefactor (see Eq.~2 in \citealt{Cooper22}), $B(r, \theta, \phi)$ is the magnetic field of the primary NS, which can be approximated by a dipole i.e. $B\approx B_\text{s}(R_\text{NS}/a)^3$ \textbf{(where $B_\text{s}$ is the surface magnetic field of the primary NS)}, and $\beta=v/c$ is the speed of the secondary NS and varies with the binary separation as

\begin{equation}
    \beta=\frac{1}{c}\sqrt{\frac{GM}{a(t)}},
\end{equation}

\noindent \textbf{where $c$ is the speed of light, $G$ is the gravitational constant, and $M$ is the primary NS mass.}

The orbit-induced electric field can accelerate particles along open magnetic field lines and move them out of the polar cap regions, creating vacuum like gaps in the magnetosphere (similar to the polar cap models of pulsar emission; \citealt{Ruderman75, Daugherty82}). We can estimate the gap height by the distance from the initial acceleration point to the point where pair production completely screens the electric field. Here, for simplicity, we assume a one-dimensional and stationary gap, and the electric field in the gap $E_\text{gap}=E_{\parallel}$. Then the gap height is \mbox{$h_\text{gap}\propto\rho^{1/2}_\text{c}B^{-1/4}E_{\parallel}^{-3/4}$}, where $\rho_\text{c}$ is the curvature radius of the magnetic field (see Eq. 19 in \citealt{Cooper22}). Assuming a fraction, $\epsilon_\text{r}$, of the acceleration power of the polar gap is converted to radio emission, we can calculate the radio luminosity,

\begin{equation}
    L_\text{r}=\epsilon_\text{r}e\Phi_\text{gap}\dot{N}=\epsilon_\text{r}eE_\text{gap}h_\text{gap}nAc,
\label{eq:L_NSs}
\end{equation}

\noindent where $e$ is the electric charge, $\Phi_\text{gap}=E_\text{gap}h_\text{gap}$ is the electric potential difference along the gap, $n=E_\text{gap}/(4\pi eh_\text{gap})$ is the charge number density in the gap, $A\approx 4\pi R_\text{NS}^2$ is the cross section of the gap, and $\dot{N}=nAc$ is the rate of accelerated particles.

With the above equations, we can calculate the radio luminosity from any point surrounding the secondary NS, which is time (or orbital separation) dependent and magnetic field line directed. In order to estimate the viewing angle dependence, following the prescription outlined in \citet{Cooper22}, we performed a numerical simulation that calculated the radio luminosity for each volume element $\Delta V$ at $(r, \theta, \phi)$ and each timestep $t$. For an observer at $(d_L, \theta, \phi)$ in the frame of the secondary NS \textbf{($d_L$ is the luminosity distance to the secondary NS)}, the observable emission is contributed by all volume elements with magnetic fields aligned with the observer (for more details about the numerical simulation see \citealt{Cooper22}). We applied this numerical simulation to obtain the viewing angle dependent emission (see below).

Figure~\ref{Pulsar_revival} shows the radio emission predicted to be produced during the final 3\,ms of the BNS inspiral, encompassing the final two orbital periods and thus two peaks of emission \citep{Cooper22}, for a range of viewing angles between $\mathbf{0^\circ}$ and $\mathbf{60^\circ}$ at an observing frequency of $\nu_\text{obs}=120$\,MHz (a plausible observing frequency for the MWA; see Section~\ref{sec:simulation}). Note that we do not expect coherent radio emission to be detectable beyond a viewing angle of $\mathbf{60^\circ}$ as no magnetic field lines are perturbed away from the background magnetic field beyond this angle (see figure 1 in \citealt{Cooper22}). We adopt the following NS parameters: a mass $M=1.4\,\text{M}_\odot$ and radius $R_\text{NS}=10^6$\,cm for both NSs, a surface magnetic field of the primary NS $B_\text{s}=10^{14}$\,G, an angle between the magnetic axis and the orbital plane $\alpha_\text{B, orb}=90^\circ$ for the primary NS, and an efficiency factor $\epsilon_\text{r}=10^{-2}$. This efficiency agrees with population studies of pulsar luminosity with voltage-like scaling and beaming models (e.g. \citealt{Arzoumanian02}). Note that the magnetic axis of the primary NS is not necessarily perpendicular to the orbital plane, and as the magnetic axis tilts towards the orbital plane the magnetic field surrounding the secondary NS can increase by a factor of 2, corresponding to a radio luminosity increase by a factor of 4 (see Eq. 25 in \citealt{Cooper22}). 
Also note that the radio luminosity scales with the magnetic field of the primary NS and the radio efficiency as $\propto(B_\text{s}/10^{14}\,\text{G})\times(\epsilon_\text{r}/10^{-2})$. In the case of a weaker magnetic field and a smaller efficiency factor, e.g. $B_\text{s}=10^{12}\,\text{G}$ and $\epsilon_\text{r}=10^{-4}$, the radio luminosity could be attenuated by a factor of $10^4$.
In Figure~\ref{Pulsar_revival}, we can see the observable fluence of the 3\,ms signal prior to the BNS merger decreases by a factor of $\sim1000$ as our line of sight moves away from the magnetic axis. At an observing angle of $\theta_\text{obs}\lesssim\mathbf{30^\circ}$ and a distance of $\lesssim150$\,Mpc, the fluence can reach $\gtrsim1000$\,Jy\,ms, which can be detected with the MWA (see Section~\ref{sec:simulation}).

\begin{figure}
\centering
\includegraphics[width=0.48\textwidth]{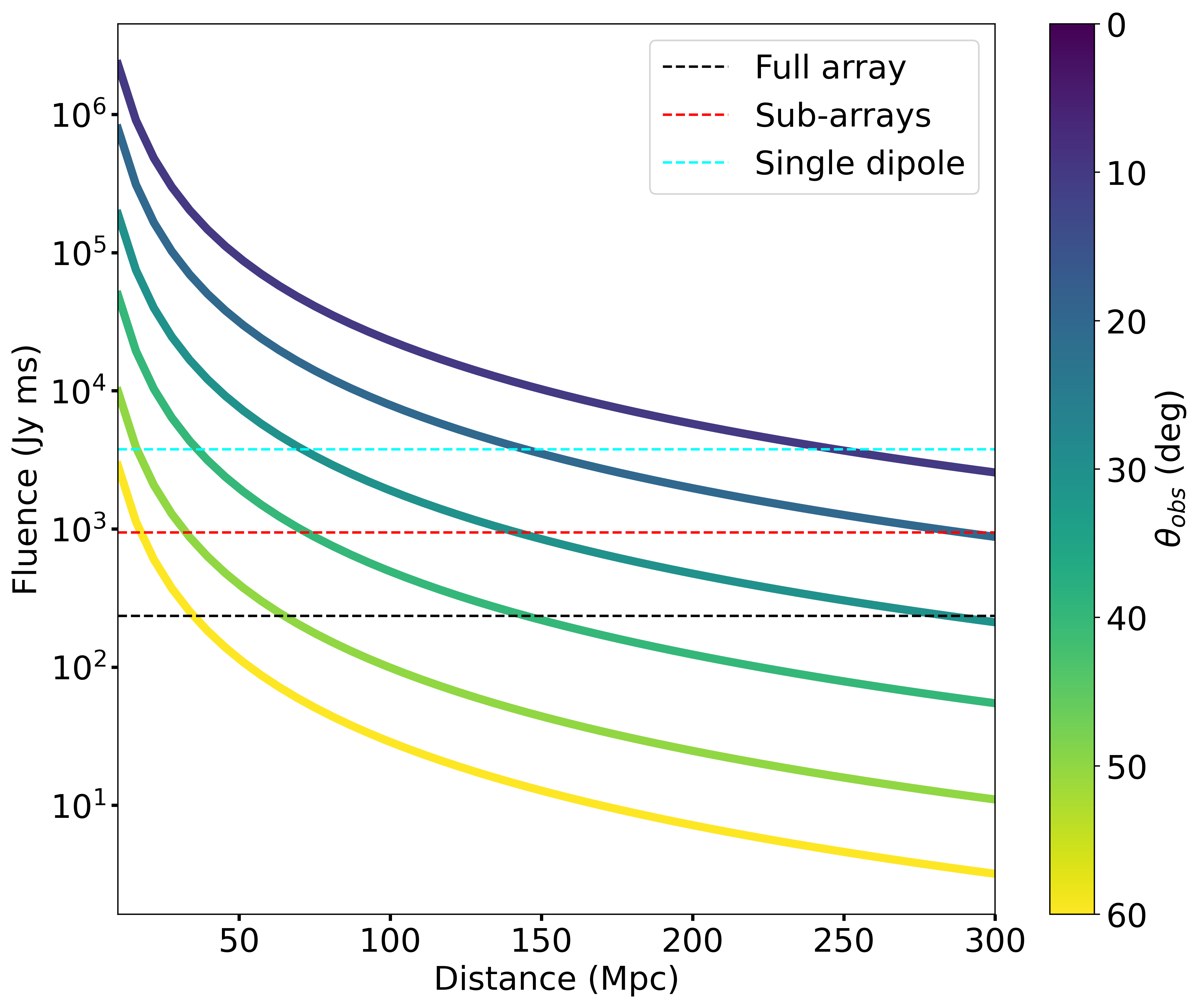}
\caption{The total fluence of the radio emission predicted to be produced during the last 3\,ms of the BNS inspiral at 120\,MHz assuming a mass $M=1.4\,\text{M}_\odot$ and radius $R_\text{NS}=10^6$\,cm for both NSs, a surface magnetic field for the primary NS $B_\text{s}=10^{14}$\,G, an angle between the magnetic axis and the orbital plane $\alpha_\text{B, orb}=90^\circ$ for the primary NS, and an efficiency factor $\epsilon_\text{r}=10^{-2}$. The solid lines represent the observable emission with the color corresponding to different viewing angles with respect to the binary merger axis based on the color bar. The three horizontal dashed lines in black, red, and cyan represent the expected sensitivity on $\sim$ms timescales of the MWA full array, four sub-arrays, and a single dipole per tile, respectively, all in the VCS mode and with incoherent beamforming (see Section~\ref{sec:simulation}).}
\label{Pulsar_revival}
\end{figure}

\subsection{Relativistic jet and ISM interaction}\label{sec:model_ism}

It has been suggested that the interaction between a Poynting flux dominated jet launched by BNS mergers and the ISM can produce a coherent radio pulse as well as prompt gamma-ray emission \citep{Usov00}. 
Given the coincident detection of GRB 170817A just 2\,s following the detection of GW170817 \citep{Abbott17, Abbott17b, Goldstein17}, we might expect this prompt radio emission to occur on similar timescales 
following BNS mergers.
The bolometric radio fluence, $\Phi_{r}$ ($\text{erg}\,\text{cm}^{-2}$), is proportional to the bolometric gamma-ray fluence 
observed from 
short GRBs, $\Phi_{\gamma}$ ($\text{erg}\,\text{cm}^{-2}$), with their ratio being $\simeq0.1\epsilon_{B}$, where $\epsilon_{B}$ is the fraction of magnetic energy in the relativistic jet \citep{Usov00}. The typical spectrum of these low-frequency waves is expected to peak at a frequency dependent on the magnetic field at the shock front

\begin{equation}
    \nu_\text{max}\simeq[0.5\,\text{--}\,1]\frac{1}{1+z}\epsilon_{B}^{1/2}\times10^6\,\text{Hz}
    \label{eq:fpeak}
\end{equation}

\noindent (in the observer's frame; \citealt{Rowlinson19b}), which is well below our observing frequency. The radio fluence at an observing frequency $\nu_\text{obs}$ is given by

\begin{equation}
    \Phi_{\nu_\text{obs}}=\frac{\beta-1}{\nu_{\text{max}}}\Phi_{r}\bigg(\frac{\nu_\text{obs}}{\nu_{\text{max}}}\bigg)^{-\beta}\,\text{erg}\,\text{cm}^{-2}\,\text{Hz}^{-1},
\end{equation}

\noindent where the spectral index is typically assumed to be $\beta=1.6$ \citep{Usov00}. Note that the bolometric radio fluence $\Phi_r$ is the fluence integrated over frequency and thus has a different unit to $\Phi_{\nu_\text{obs}}$.

We can predict the fluence of the coherent radio emission produced during the BNS merger using the above equations. The gamma-ray fluence may be inferred using

\begin{equation}
    \Phi_{\gamma}=\frac{(1+z)\,E_{\gamma,\text{iso}}}{4\pi d_L^2}\,\text{erg}\,\text{cm}^{-2},
\end{equation}

\noindent where $E_{\gamma,\text{iso}}$ represents the isotropic-equivalent gamma-ray energy, and ranges between $(0.04\text{--}45)\times10^{51}$\,erg with a median value of $1.8\times10^{51}$\,erg (inferred from a population of short GRBs; \citealt{Fong15}). However, the above calculation applies only to an on-axis jet, i.e. the relativistic jet points along our line-of-sight, which is a reasonable assumption in searching for radio counterparts to GRBs (e.g. \citealt{Rowlinson19b, Rowlinson21, Anderson20, Tian22a, Tian22b}). In the case of GW detections, the relativistic jet launched by the BNS merger is likely to point away from the Earth, resulting in no GRB detection. Therefore, for GW triggers with MWA it is necessary to consider the attenuation of the predicted emission with the viewing angle
(see Section~\ref{sec:simulation}). Note that for discussions in this paper we assume the relativistic jet aligns with the orbital angular momentum of the binary system (e.g. \citealt{Abbott17b}).

In order to calculate the viewing angle-dependent radio emission, we assume a structured jet model i.e. the angular distribution of kinetic energy within the jet, which may arise from the central engine activity and/or the interaction of the jet with the ISM \citep{Gottlieb18, Lazzati18, Xie18}. There are several variants of the structured jet models, including a top-hat jet \citep{Donaghy06}, a power-law jet \citep{Dobie20}, or a Gaussian jet \citep{Resmi18}. Given that current observations of GRBs do not allow us to distinguish between these different jet structures and that much evidence appears to support Gaussian structured jets for GRBs (e.g. \citealt{Lamb18, Lamb19, Howell19, Cunningham20}),
here we adopt a Gaussian jet model where the distribution of kinetic energy and Lorentz factor within the jet is given by

\begin{equation}
    E(\theta)=E_\text{iso}\,e^{-(\theta/\theta_0)^2},\,\,\text{and}
\end{equation}

\begin{equation}
    \Gamma(\theta)=1+(\Gamma_0-1)\,e^{-(\theta/\theta_0)^2},
\end{equation}

\noindent where $\theta$ is the polar angle from the jet's axis, $\theta_0$ is the angular scale of the jet opening angle, and $E_\text{iso}$ and $\Gamma_0$ are the isotropic-equivalent energy and Lorentz factor of the jet's core, respectively. There are different methods of constraining the jet opening angle. While observations of jet breaks in short GRB afterglows suggest a typical jet opening angle of $\mathbf{16^\circ\pm10^\circ}$ \citep{Fong15}, a comparison between the rates of BNS mergers and short GRBs points to highly collimated GRB jets with opening angles $\approx\mathbf{6^\circ}$ \citep{Beniamini19}. Note that the latter constraint on the jet opening angle was improved in \citet{Sarin22} to $\approx\mathbf{15^\circ}$. Here for completeness we consider the model emission under both a narrow ($\theta_0=\mathbf{6^\circ}$) and wide ($\theta_0=\mathbf{16^\circ}$) jet.

The jet emission viewed off-axis may be calculated as follows. Assuming a Lorentz factor of $\Gamma_0\sim1000$ (e.g., \citealt{Hotokezaka19, Dobie20}), we have the relativistic beaming cone of emitters $1/\Gamma\ll\theta_0$. In this case, the observed radio emission scales with the on-axis emission as

    \[
 \frac{\Phi_{\nu_\text{obs}}(\theta)}{\Phi_{\nu_\text{obs}}(\theta_0)} = 
  \begin{cases} 
   \frac{E(\theta)}{E(\theta_0)} & \text{} \theta < \theta_0 \\
   \text{max}\left[\frac{E(\theta)}{E(\theta_0)}, q^{-4}\right] & \text{} \theta_0 < \theta < 2\theta_0 \\
   \text{max}\left[\frac{E(\theta)}{E(\theta_0)}, q^{-6}(\theta_0\Gamma)^2\right] & \text{} \theta > 2\theta_0
  \end{cases}
\]

\noindent where the term in each row containing $q=(\theta-\theta_0)\Gamma$ represents the case when `off line-of-sight' emitters (i.e., the angle between the velocity of emitters and our line-of-sight is larger than $1/\Gamma$) become dominant \citep{Beniamini19, Beniamini19b}.

\begin{figure}
\centering
\includegraphics[width=0.5\textwidth]{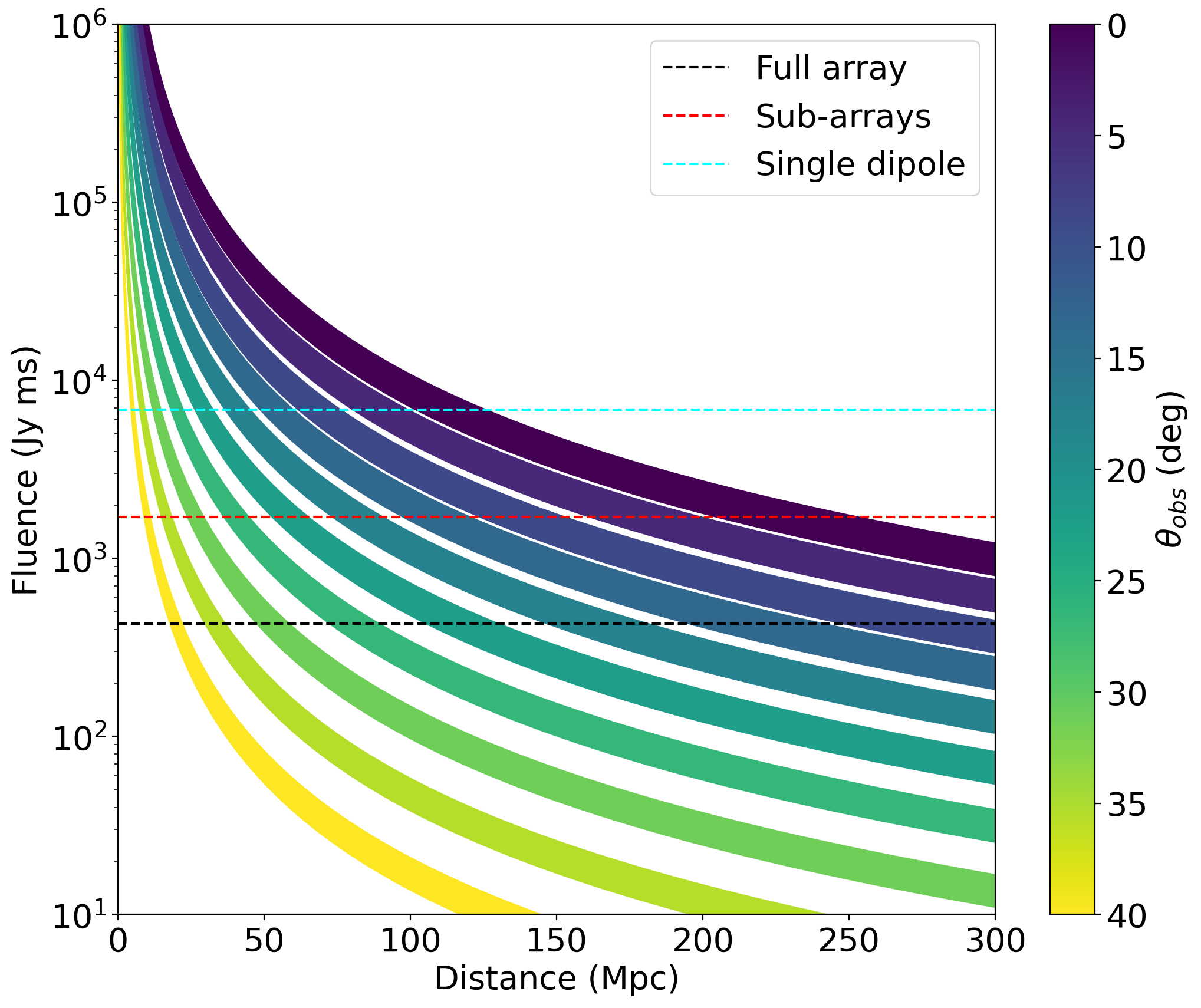}
\caption{The fluence of the prompt radio signal predicted to be produced by the relativistic jet and ISM interaction at 120\,MHz assuming a Gaussian jet with an angular scale of $\mathbf{16^\circ}$ (see Section~\ref{sec:model_ism}). The regions in different colors show the radio fluence predictions corresponding to different viewing angles from $\mathbf{0^\circ}$ (on-axis) to $\mathbf{40^\circ}$ (off-axis), with the uncertainties (depicted by the width of the different color regions) resulting from the peak frequency of the prompt radio emission at the shock front (see Eq. \ref{eq:fpeak}). The three horizontal dashed lines are the same as for Figure~\ref{Pulsar_revival}.} 
\label{GRB_jet_16deg}
\end{figure}

Figure~\ref{GRB_jet_16deg} shows the radio emission predicted to be produced by the relativistic jet-ISM interaction for a range of viewing angles between $\mathbf{0^\circ}$ (on-axis) and $\mathbf{40^\circ}$ (off-axis) at an observing frequency of $\nu_\text{obs}=120$\,MHz. We adopt the following jet parameters: $E_{\text{iso}}=1.8\times10^{51}$\,erg; $\epsilon_{B}=10^{-2}$; $\theta_0=\mathbf{16^\circ}$; and $\Gamma_0=1000$ \citep{Fong15}. We can see the observable model emission drops with viewing angle. At a distance of 200\,Mpc, while the on-axis radio fluence can reach $>1000$\,Jy\,ms, the off-axis fluence for $\theta_\text{obs}=\mathbf{40^\circ}$ drops to below 10\,Jy\,ms, which means the detectability of this emission model is largely determined by our viewing angle (see Section~\ref{sec:simulation}). We note that in the case of a narrow jet with $\theta_0=\mathbf{6^\circ}$, the decrease of fluence with viewing angle is more significant, with a viewing angle of $\mathbf{(6^\circ/16^\circ)\times40^\circ=15^\circ}$ resulting in 
a predicted radio fluence below 10\,Jy\,ms (see Figure~\ref{GRB_jet_6deg} in Appendix~\ref{appendix:GRB_jet_6deg}).

\subsection{Persistent pulsar emission}\label{sec:model_pers}

If the merger remnant is a magnetar, we may expect there to be persistent radio emission powered by dipole magnetic braking during the lifetime of the magnetar \citep{Totani13, Metzger17}. 
The duration of this emission is largely uncertain due to the unknown equation of state and lifetime of the magnetar remnant. However, assuming the plateau phase observed in the X-ray afterglow of short GRBs is powered by the magnetar remnant and its ending is due to the magnetar collapse (see Section~\ref{sec:model_col}), we might expect this persistent radio emission to last until $\sim1000$--10000\,s post-merger \citep{Tang19, Sarin20}. Note that in the case of an extremely low binary mass (i.e. $\lesssim M_{\text{max}}$, the maximum mass of stable NSs; \citealt{Lattimer01}), the magnetar remnant might be indefinitely stable and would therefore not collapse (e.g., \citealt{Bucciantini12, Giacomazzo13}).
The luminosity of this emission is given by \citep{Pshirkov10},

\begin{equation}
    L=\epsilon_r\,\dot{E},
\end{equation}

\noindent where $\epsilon_r$ is the radio emission efficiency and $\dot{E}$ is the standard pulsar spin-down luminosity \citep{Zhang01},

\begin{equation}
    \dot{E}=\frac{16\pi^4}{3}\frac{B^2 R^6}{P^4 c^3}\,\text{sin}^2\alpha,
\end{equation}

\noindent where $P$, $B$, $R$, and $\alpha$ are the spin period, surface magnetic field, radius, and magnetic inclination of the magnetar remnant, respectively, and $c$ is the speed of light. Note that the above expression assumes a braking index of 3 for the magnetar, which usually differs from measured values of millisecond magnetars \citep{Lasky17, Sasmaz19}. If we take into account the beaming fraction $\Omega/(4\pi)$ of the radio emission, the detectable flux density is given by

\begin{equation}
    F_{\nu_\text{obs}}=\frac{(1+z)\,L}{\Omega\,\nu_\text{obs}\,d_L^2}.
    \label{eq:pulsar}
\end{equation}

We assume the same radio emission efficiency as in Section~\ref{sec:model_NSB} i.e. $\epsilon_\text{r}=10^{-2}$.
Two main sources of uncertainty in the predicted flux density are: the magnetic inclination angle of the magnetar remnant $\alpha$ (due to the unknown physics of the NS magnetic field and equation of state; \citealt{Cutler02}) and the beaming fraction of the radio emission $\Omega/(4\pi)$ (due to the unknown physics of the pulsar radio emission; \citealt{Kalogera01}).
As the magnetic pole of a NS is expected to align with the spin axis at the birth time and become misaligned with time (the orthogonalisation timescale due to bulk viscosity inside a NS is largely uncertain depending on the NS spin frequency, magnetic field strength, and temperature, and could be as short as seconds; \citealt{Dall09, Lander18}), here we adopt a fiducial value of $\mathbf{30^\circ}$ for $\alpha$. For the beaming fraction we consider a range of $0.01<\Omega/(4\pi)<1$ (e.g. \citealt{gourdji20}). Note that here the beaming fraction is for the off-axis viewing angle consideration rather than being physical i.e. the observed flux density is given by Eq.~\ref{eq:pulsar} if the impact angle of our line of sight to the magnetic axis is within the solid angle $\Omega$ and zero otherwise.

\begin{figure}
\centering
\includegraphics[width=0.5\textwidth]{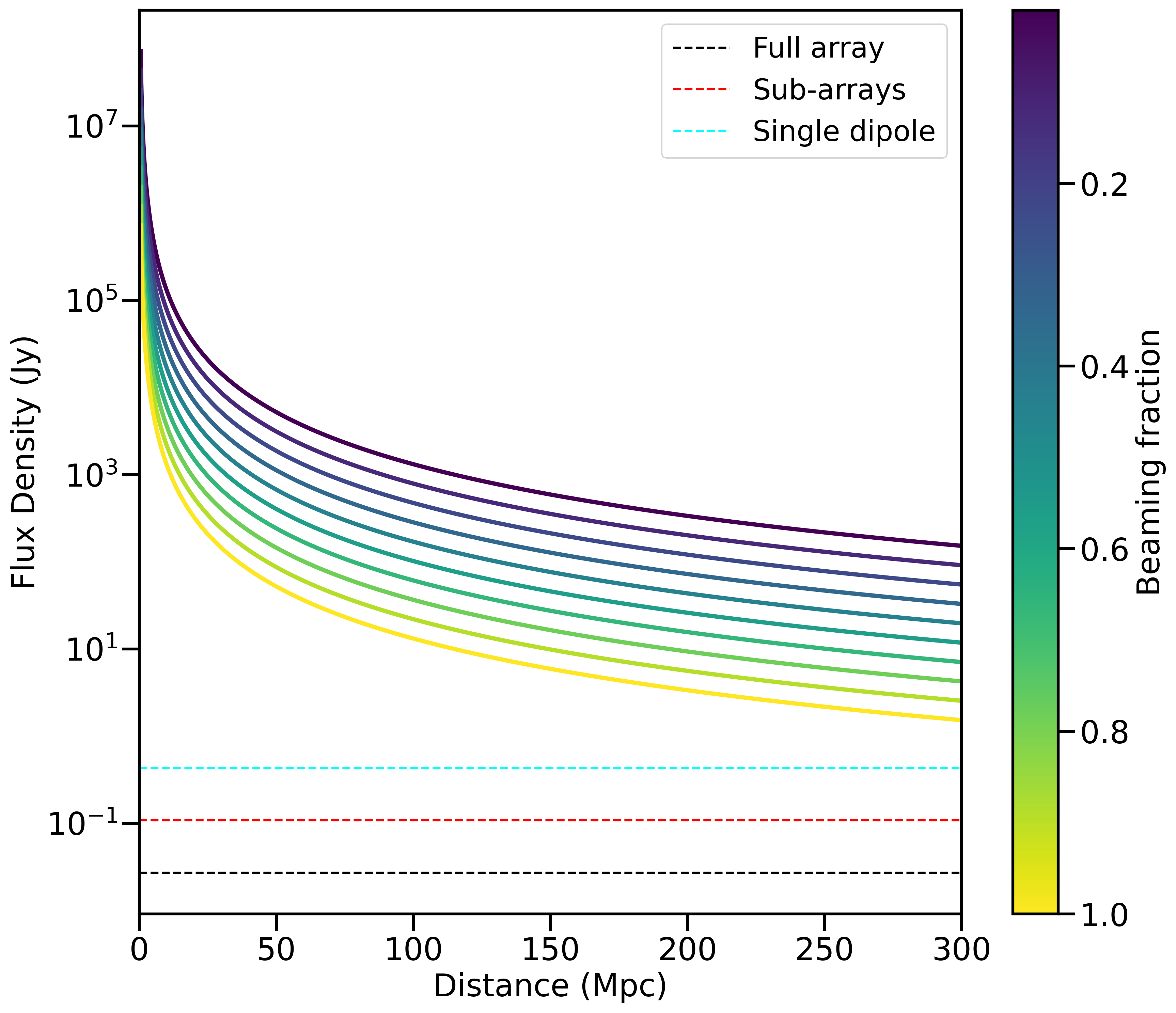}
\caption{The predicted flux density for the persistent radio emission from the dipole radiation of a magnetar remnant at 120\,MHz (see Section~\ref{sec:model_pers}). We assumed a radio emission efficiency of $\epsilon_\text{r}=10^{-2}$ and a fiducial angle of $\mathbf{30^\circ}$ for the magnetic inclination of the magnetar. The solid lines in different colors represent the observable emission from a low luminosity magnetar (i.e. $B=8\times10^{15}$\,G and $P=30$\,ms; see Section~\ref{sec:model_pers}) for different beaming fractions. The horizontal dashed lines in black, red, and cyan represent the expected sensitivity on 30\,min timescales of the MWA full array, four sub-arrays and a single dipole per tile, respectively, all in the standard correlator mode (see Section~\ref{sec:simulation}).}
\label{pulsar_emission}
\end{figure}

Figure~\ref{pulsar_emission} shows the predicted persistent radio emission from the magnetar remnant formed by BNS mergers in the case of the radiation beam pointing towards us for a range of beaming fractions at an observing frequency of $\nu_\text{obs}=120$\,MHz (see Section~\ref{sec:simulation}). We adopt magnetar parameters of $B=8\times10^{15}$\,G and $P=30$\,ms, corresponding to a low luminosity magnetar given the distribution of magnetar parameters derived from a population of short GRBs (see figure 8 in \citealt{Rowlinson19}), and note that even in the case of a smaller radio emission efficiency (e.g. $\epsilon_\text{r}=10^{-4}$; \citealt{Szary14}) the predicted persistent radio emission would still be bright enough to be detected by the MWA. 

\subsection{Magnetar collapse}\label{sec:model_col}

\textbf{If the magnetar remnant is supramassive,}
it will collapse into a BH inevitably, ejecting its magnetosphere and possibly producing a short burst of coherent radio emission \citep{Falcke14, Zhang14}. 
Given the timescale of the magnetar collapse inferred from the X-ray afterglow of short GRBs (see Section~\ref{sec:model_pers}), we might expect this radio emission to occur $\sim$1000--10000\,s post-merger.
Assuming a fraction $\epsilon$ of the magnetic energy in the magnetar's magnetosphere $E_B$ is converted into the radio emission, we can write the bolometric radio fluence as

\begin{equation}
    \Phi_r=\epsilon\,E_B=\epsilon\,\frac{B^2R^3}{6}.
\end{equation}

\noindent Taking into account the beaming of the radio emission as in Section~\ref{sec:model_pers}, we can show that the observable radio fluence is

\begin{equation}
    \Phi_{\nu_\text{obs}}=\frac{(1+z)\,\Phi_r}{\Omega\,\nu_\text{obs}\,d_L^2}.
\end{equation}

\noindent Note that the above equation applies only if our line of sight falls in the radiation beam, as noted in Section~\ref{sec:model_pers}.

\begin{figure}
\centering
\includegraphics[width=0.5\textwidth]{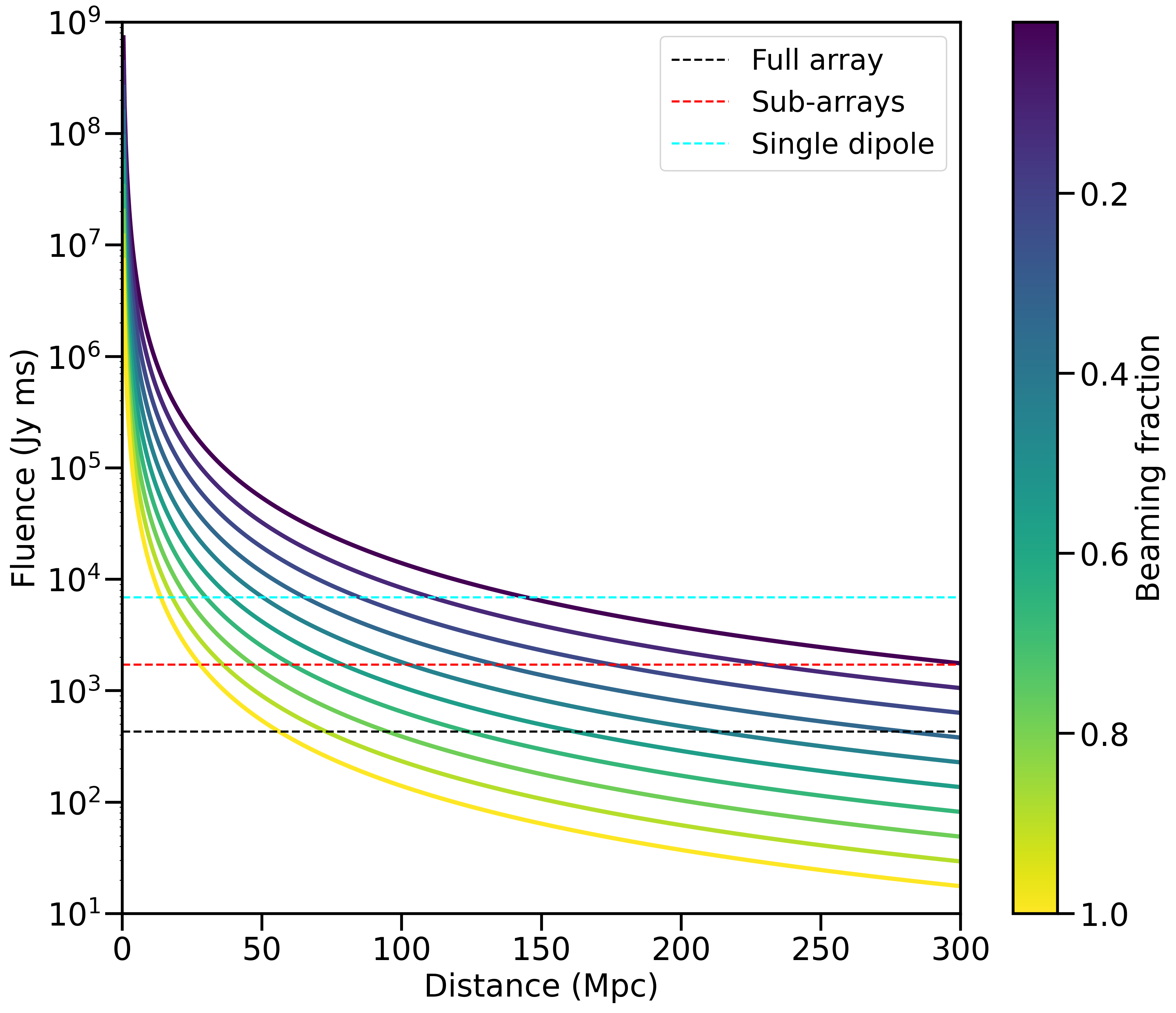}
\caption{The predicted fluence for the radio burst produced during the collapse of the magnetar remnant at 120\,MHz (see Section~\ref{sec:model_col}). We assumed a magnetic energy conversion efficiency of $\epsilon=10^{-6}$. The solid lines in different colors represent the observable emission resulting from the collapse of a typical magnetar remnant (i.e. $B=2\times10^{16}$\,G; see Section~\ref{sec:model_col}) for different beaming fractions. The three horizontal dashed lines are the same as for Figure~\ref{Pulsar_revival}.}
\label{collapse}
\end{figure}

Figure~\ref{collapse} shows the predicted radio burst resulting from the collapse of the magnetar remnant in the case of the radiation beam pointing towards us for a range of beaming fractions $0.01<\Omega/(4\pi)<1$ at an observing frequency of $\nu_\text{obs}=120$\,MHz (see Section~\ref{sec:simulation}). We assume an efficiency of converting magnetic energy into radio emission of $\epsilon=10^{-6}$ (upper limit suggested by, e.g. \citealt{Rowlinson21}), and a typical magnetar remnant (i.e. $B=2\times10^{16}$\,G; \citealt{Rowlinson19}). 
We can see the predicted emission varies with beaming fraction by more than two orders of magnitude from $\gtrsim100$\,Jy\,ms at $\Omega/(4\pi)=1$ to $\gtrsim10000$\,Jy\,ms at $\Omega/(4\pi)=0.01$. Therefore, the detectability of this model emission is dependent on both beaming fraction and viewing angle (see Section~\ref{sec:simulation}).

\section{A population study for the radio counterparts to GW sources}\label{sec:simulation}

In this Section, we perform a population study for the radio counterparts to GW sources in the context of the four coherent emission models described in Section~\ref{sec:models} in order to access the viability of MWA detecting these signals in dedicated triggered follow-up during O4 and beyond.
In order to detect prompt radio emission from BNS mergers we need to consider the sky coverage for the predicted GW detections in O4 (see Figure~\ref{GW_map}) as well as the viewing angle dependence of the emission models (see Section~\ref{sec:models}).
We use a Monte Carlo method for simulating $10^7$ binary systems with random inclinations and distances within the LVK O4 horizon 
(i.e. the inclination between the orbital angular momentum of the binary and the line of sight). Then we apply a GW detection criterion to determine the population of BNS mergers likely to be detected by LVK. Assuming the same intrinsic radio emission for all BNS mergers as derived in Section~\ref{sec:models}, we can calculate the observable radio fluence or flux density (depending on the source distance and inclination angle) for each simulated GW detection, and compare it to the instrument sensitivity for determining the LVK BNS merger GW-radio jointly detectable fraction with the MWA.

\subsection{GW detection criterion}\label{sec:GW_criterion}

The detectability of a GW inspiral signal by a LIGO-Virgo type interferometer depends on the property of the binary system as well as the sensitivity profile of the interferometer \citep{Finn93}. A full analysis requires the consideration of the chirp mass of the binary, the luminosity distance to the binary, and the binary localisation and orientation in the frame of the interferometer. Here, we need to consider only two parameters, the luminosity distance $d_L$ and the inclination angle $\theta_\text{obs}$, as these are the parameters that determine the predicted radio emission from BNS mergers (see Section~\ref{sec:models}). Then the GW detection criterion for the LVK network assuming a S/N threshold of eight is given by \citep{Duque19}

\begin{equation}
    d_L<H(\theta_\text{obs})=\sqrt{\frac{1+6\cos^2{\theta_\text{obs}}+\cos^4{\theta_\text{obs}}}{8}}\,\overline{H},
\end{equation}

\noindent where $\overline{H}$ is the sky position averaged horizon (190\,Mpc for $1.4+1.4\,\text{M}_\odot$ BNS systems during the O4 run; \citealt{Abbott20}). We note that the early phase of O4 has a sensitivity close to O3 and an actual horizon limit of $\sim160$\,Mpc. However, the simulation results below remain the same for different horizon limits.

We started our population study by simulating $10^7$ BNS systems that are homogeneously distributed within the horizon ($H=\sqrt{\frac{5}{2}}\,\overline{H}\approx300\,\text{Mpc}$) and have isotropically distributed binary angular momentum directions. Applying the above criterion, we obtained $\sim29\%$ detected by the GW interferometer network, with their distribution in inclination angle shown in Figure~\ref{GW_angle} (black dotted line). We can see the mean inclination angle of the GW detected events is $\sim38^\circ$, which is consistent with previous works (e.g. \citealt{Duque19, Mochkovitch21}). This fraction of GW detected events were further filtered with a radio detection criterion for determining the GW-radio jointly detectable BNS merger population (see Section~\ref{sec:radio_criterion}).

\subsection{Radio detection criterion}\label{sec:radio_criterion}

For this analysis, we assume that a GW source is detectable in the radio band as long as its observable radio emission, as determined by the models in Section~\ref{sec:models}, is above the sensitivity of the radio telescope used for follow-up. Note that this criterion is necessary but not sufficient for a real detection, which also depends on follow-up time and the arrival of radio signals. Here, for simplicity we assume that the MWA is capable of capturing all the four model emissions presented in Section~\ref{sec:models} regardless of their arrival times (for more discussion see Section~\ref{sec:strategies}).

We chose to test radio detections at 120\,MHz and 200\,MHz for a balance between sky coverage and detection sensitivity. The MWA has a larger FoV at lower frequencies, which is more ideal for covering the GW positional uncertainties 
as shown in Table~\ref{mode_summary}. However, considering the model emission presented in Section~\ref{sec:models} could potentially be FRB-like, and the fact that most FRB signals have been detected at $>300$\,MHz (e.g. \citealt{Chawla20, Pilia20, Parent20, CHIME21}), we might expect 
a higher chance of detecting coherent radio counterparts to GWs at higher observing frequencies.
As a compromise, in this paper all properties of the MWA including the FoV and the sensitivity are quoted for 120\,MHz and 200\,MHz (see Table~\ref{mode_summary}). Note that while the MWA has the optimal sensitivity at 150\,MHz, 120\,MHz will provide a larger FoV in an RFI-quiet part of the MWA band while gaining additional dispersion delay and therefore time for getting on-target. Therefore, we chose an observing frequency of 120\,MHz, which is on the lower frequency end of FRB detections \citep{Pleunis21}.



As previously mentioned, the earliest LVK GW alerts of BNS mergers will have poor positional localisations, however, most of the models discussed in Section~\ref{sec:models} strongly motivate the need for MWA to be on target during, if not before the merger. 
An exciting addition to the O4 public alerts is the Early-Warning Alerts from pipelines capable of detecting GWs from the inspiral before the merger of a binary with at least one NS component: GstLAL \citep{Cannon12, Sachdev20}, MBTAOnline \citep{Adams16},  PyCBC Live \citep{Nitz18, Canton21}, and SPIIR \citep{Chu22}.
However, such alerts will not contain any positional information.\footnote{\url{https://emfollow.docs.ligo.org/userguide/early\_warning.html}}
Rather than waiting for an accurate sky position, we instead need to configure the MWA to observe as much of the sky as possible on receiving a GW alert while also taking advantage of the telescope's fortuitous position under one of the two highest sensitivity sky regions of the LVK network (see Figure~\ref{GW_map}).
In order to further increase our chances of a successful detection at early times, 
we experimented with three different ways of configuring the MWA that would 
test for the best compromise between sky coverage and sensitivity (see also Figure~\ref{prob_map} and Table~\ref{mode_summary}): 

\begin{enumerate}
    \item The full array (128 tiles $\times$ 16 dipoles) with a single primary beam that is centered on the position of the highest sensitivity  region of the LVK network over the Indian Ocean (see Figure~\ref{GW_map} and Figure~\ref{prob_map}a);
    
    \item A single dipole per tile, which provides a very widefield Zenith pointing (see Figure~\ref{prob_map}a); and
    
    \item Splitting the full array into four sub-arrays of 32 tiles each, creating four overlapping primary beams that tile the highest LVK network  sensitivity region, overlapping at 50\% power (see Figure~\ref{prob_map} for the different MWA beam tiling configurations that we tested).

\end{enumerate}

In the case of option 3, we further trialled four different sub-array beam pointing configurations to maximise our coverage of the highest sensitivity region of the LVK network. Specifically, one beam is always centred on the highest sensitivity GW region 
with the other three beams overlapping at their 50\% power 
(the different pointing configurations are listed in Table~\ref{MWA_pointings} and depicted in Figure~\ref{prob_map}). We then tested each of these beam tiling configurations for which would provide the highest probability of detecting coherent radio emission from a BNS merger 
(see Section~\ref{sec:GW_map}).

In Table~\ref{mode_summary}, we list the approximate MWA sensitivity for the three observing modes 
on timescales of 1\,ms (assuming an MWA VCS observation and incoherent beamforming) and 30\,min  (assuming a standard MWA correlator observation). 
Note that the quoted sensitivities for the single dipole and the sub-array modes
are estimated by simply assuming that the sensitivity for the full array pointed at the zenith scales with the number of dipoles in use.  
The sensitivity on 1\,ms timescales is appropriate for considering the detectability of prompt radio emission predicted to be produced by the NS magnetosphere interaction (see Section~\ref{sec:model_NSB}), jet-ISM interaction (see Section~\ref{sec:model_ism}), and magnetar remnant collapse (see Section~\ref{sec:model_col}), and the sensitivity on 30\,min timescales is for persistent radio emission produced by the magnetar remnant (see Section~\ref{sec:model_pers}). These sensitivities, combined with the GW detection criterion, form our criterion for the joint detection of simulated BNS mergers
(for discussion on our observing strategies see Section~\ref{sec:strategies}). 

\begin{table}
\centering
\caption{A summary of the three observing modes (see Section~\ref{sec:radio_criterion}), including the field of view (down to 20\% of the primary beam) at both 120\,MHz and 200\,MHz, and the sensitivity for 1\,ms and 30\,min integrations. Here the sensitivity is quoted for 185\,MHz (extensively used for MWA GRB triggered follow-up; \citealt{Anderson20, Tian22a, Tian22b}), which we expect to be accurate to within $\sim30\%$ at 120\,MHz and 200\,MHz \citep[note that the MWA sensitivity is extremely dependent on the sky position and observational elevation,][]{Sokolowski17}.}
\resizebox{\columnwidth}{!}{\vspace{-0cm}\begin{tabular}{l l l l l}
\hline
Observing mode & \multicolumn{2}{c}{Field of view ($\text{deg}^2$)} & \multicolumn{2}{c}{Sensitivity (Jy)} \\
\cline{2-5} & 120\,MHz & 200\,MHz & 1\,ms & 30\,min\\
\hline
Full array & 990 & 269 & 136 & 0.027 \\
Single dipole per tile & 4838 & 3196 & 2172 & 0.432 \\
Four sub-arrays & 3297 & 896 & 543 & 0.108 \\
\hline
\end{tabular}}
\label{mode_summary}
\end{table}

\subsection{GW-radio jointly detectable population}\label{sec:GW_radio_joint}


Using the simulated population of BNS mergers described in Section~\ref{sec:GW_criterion}, from which we expect the LVK to detect $\sim29\%$ within the O4 horizon, we now use the models described in Section~\ref{sec:models} to  
estimate the fraction of  events that could be detected with the MWA. 
We assume all BNS mergers produce coherent radio emission described by all four models in Section~\ref{sec:models}, and we adopt 
the model parameters as described  
unless otherwise stated. 
Here we assume all GW sources are located in the MWA field of view and can be detected by the MWA as long as their predicted radio emission is above the MWA sensitivities given in Table~\ref{mode_summary}.
The LVK sky sensitivity to GW events
projected for O4 (see Figure~\ref{GW_map}) and variations in the MWA sensitivity due to the beam response and observational elevation will be considered in Section~\ref{sec:GW_map}.

\begin{figure*}
\centering
\includegraphics[width=\textwidth]{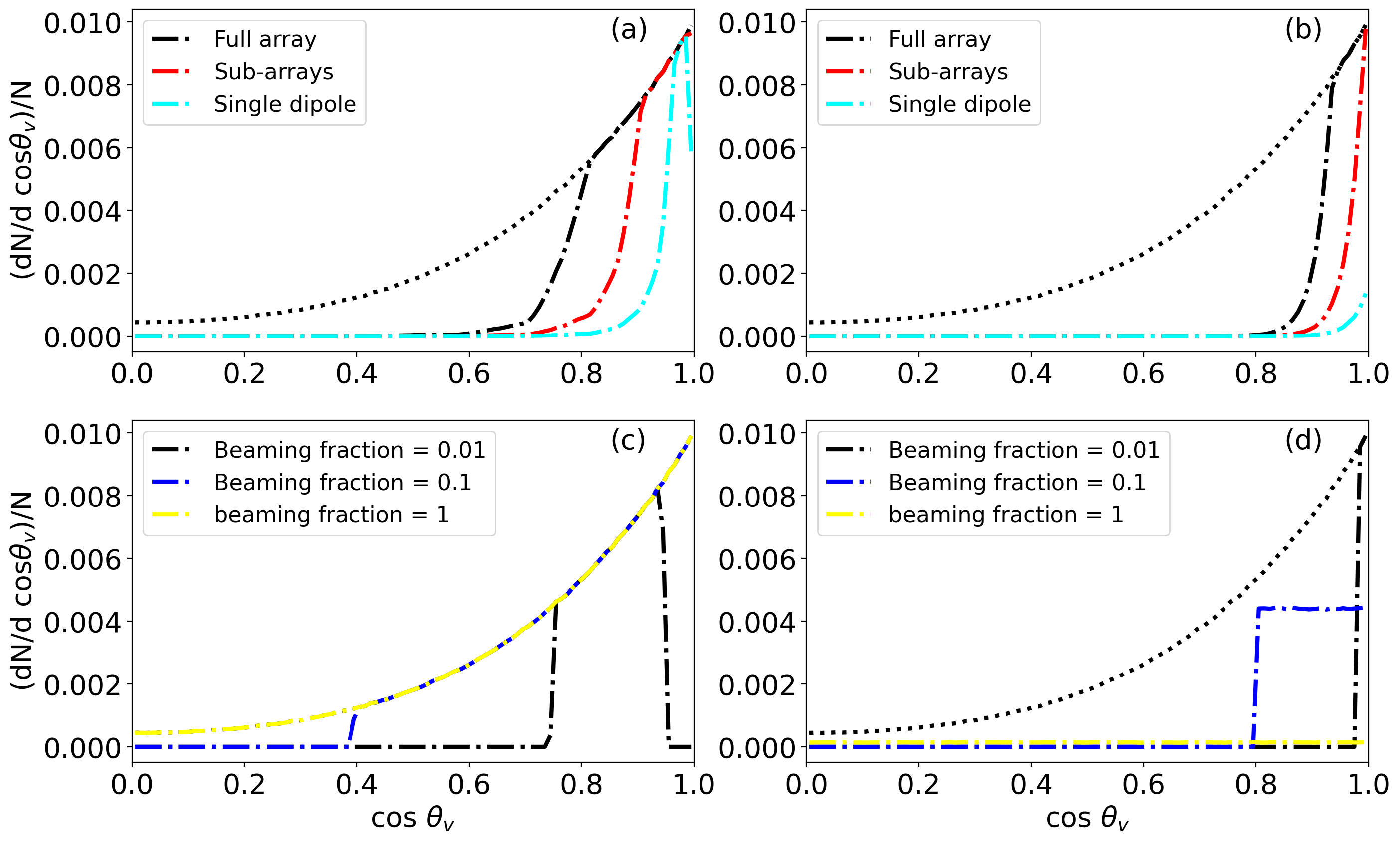}
\caption{Differential distributions as a function of inclination angle of GW detected events in the simulated BNS population (black dotted line) and GW-radio jointly detectable events (dashdot lines) for the four coherent emission models introduced in Section~\ref{sec:models}. (a) The interaction of NS magnetospheres. The dashdot lines represent those events with radio emission predicted by the NS interaction model to be detectable by the MWA with the black, red, and cyan corresponding to detections by the full array, sub-arrays and single dipole (see Section~\ref{sec:radio_criterion}). (b) The jet - ISM interaction. Here we show the distribution of GW events with radio fluence predicted by the jet-ISM interaction model to be above the MWA sensitivities (assuming a 10\,ms pulse). (c) The persistent pulsar emission from the magnetar remnant. Given the predicted emission is so bright that its detectability is only dependent on the viewing angle (see Section~\ref{sec:model_pers}), here we show the distribution of radio detectable events for the three beaming fractions, with the dashdot lines in black, blue, and yellow representing those events with a pulsar beaming fraction of 0.01, 0.1, and 1, respectively. Note that the black dotted line representing the GW detected population overlaps with the yellow dashdot line (see Section~\ref{sec:GW_radio_joint}). 
(d) The magnetar collapse. Here we show the distribution of GW events with radio emission predicted by the magnetar collapse model to be detectable by the MWA full array (assuming a 10\,ms pulse).}
\label{GW_angle}
\end{figure*}

In Figure~\ref{GW_angle}, we display the detectable fraction of coherent radio emission from BNS mergers for the four different models described in Section~\ref{sec:models} 
as a function of merger inclination angle. In each subplot, the LVK-detectable BNS population is shown as a dotted black curve. The detectable fraction of radio emission for each model using the different observing modes (described in Section~\ref{sec:radio_criterion}) or assuming different beaming fractions are shown as coloured curves. 

For the NS interaction model (see Section~\ref{sec:model_NSB} and panel (a) of Figure~\ref{GW_angle}), 
we assumed a pulse width of 3\,ms and no scattering, and converted the sensitivity from a flux density limit (which scales as $t^{-1/2}$) to a fluence limit using
\begin{equation}
    \text{Fluence}=\text{Flux}\times(\text{width}/1\,\text{ms})^{1/2}\,\text{Jy\,ms}.
    \label{eq:sensi}
\end{equation}

\noindent The fractions of GW-radio joint detections by the MWA full array, sub-arrays, and single dipole with respect to the LVK O4 detectable population are 59\%, 38\%, and 18\%, respectively.
\textbf{Note that the detectable fraction drops 
as we approach a viewing angle of 
$\cos{\theta_v}=1.0$ for the single dipole, which can be attributed to a drop in the predicted radio fluence for NS interactions when $\theta_v<10^\circ$ (see Section 3.5 in \citealt{Cooper22}) and the lower sensitivity of the single dipole compared to the other two observing modes.} 

Similarly, for the jet-ISM interaction model (see Section~\ref{sec:model_ism}) we plot the joint event detections by the different observing modes of the MWA in panel (b) of Figure~\ref{GW_angle}. Here we assume a pulse width of 10\,ms for our sensitivity estimate using Eq.~\ref{eq:sensi}, which, in the absence of detected prompt radio emission from BNS mergers, is based on known rest-frame intrinsic durations of FRBs with known redshifts and no scattering features \citep{Hashimoto19, Hashimoto20b}.
The detection fractions for the prompt radio emission produced by the jet-ISM interaction by the MWA full array, sub-arrays, and single dipole in the LVK O4 detectable population are 27\%, 11\%, and 1\%, respectively.


For the persistent pulsar emission model, as demonstrated in Section~\ref{sec:model_pers}, the MWA in any observing mode can detect the predicted emission up to the O4 horizon as long as the radiation beam of the magnetar remnant points towards us. We show the population with persistent radio emission detectable by the MWA as a function of inclination angle for three different beaming fractions (dashdot lines in different colors) in panel (c) of Figure~\ref{GW_angle}. In the case of isotropic pulsar emission (i.e. beaming fraction = 1) the MWA can detect all GW detected events, as shown by the overlapping of the black dotted and yellow dashdot lines. The detectable fractions for beaming fractions of 0.1 and 0.01 are 91\% and 43\%, respectively. There is a drop in radio detectable events around $\cos{\theta_\text{v}}\approx0.95$ for the beaming fraction $=0.01$ (black dashdot line). This can be attributed to our assumption of the magnetic inclination $=\mathbf{30^\circ}$ for the magnetar remnant (a major source of uncertainty; see Section~\ref{sec:model_pers}), which means if our line of sight aligns with the spin axis of the magnetar (i.e. small $\theta_\text{v}$), the radiation beam along the magnetic axis will point away from us.


Similarly, for the magnetar collapse model (see Section~\ref{sec:model_col}) we plot the distribution of radio detectable events for three different beaming fractions in panel (d) of Figure~\ref{GW_angle}. As shown in Figure~\ref{collapse}, the MWA can only detect the predicted emission up to a certain distance depending on the beaming fraction (different from the persistent pulsar emission model due to the faintness of the predicted emission). 
For this model, we only show the detectable fraction if using the full MWA array (128 tiles with a single primary beam). This is reasonble given the magnetar collapse is likely to occur minutes to hours following the BNS merger when we are likely to have better positional information 
(for a comparison between different observing modes see Section~\ref{sec:GW_map}). The fractions of detectable events for beaming fractions of 0.01, 0.1, and 1 are 7\%, 30\%, and 5\%, respectively. Note that the detectable fraction peaks at a beaming factor of $\sim0.1$, which is due 
to a balance between the brightness of the emission (i.e. the maximum distance we are sensitive to and the number of GW sources within that volume) and the probability of the emission beam encompassing our line of sight.


\subsection{Radio detectability rate of GW events by MWA}\label{sec:GW_map}

Apart from the intrinsic brightness of coherent radio emission and the viewing geometry of BNS mergers (including inclination angle and distance), the radio detectability rate of GW events also depends on the GW localisation in the frame of the telescope.
Taking the three MWA observing modes with different sensitivities and sky coverages, i.e. the full array, single dipole, and sub-arrays (see Section~\ref{sec:radio_criterion}), we can estimate the radio detectability rates based on the sensitivity map of the LVK detector network, which will provide a metric for assessing the success rate of our observing strategies (see Section~\ref{sec:strategies}).


\begin{figure*}
\centering
\includegraphics[width=\textwidth]{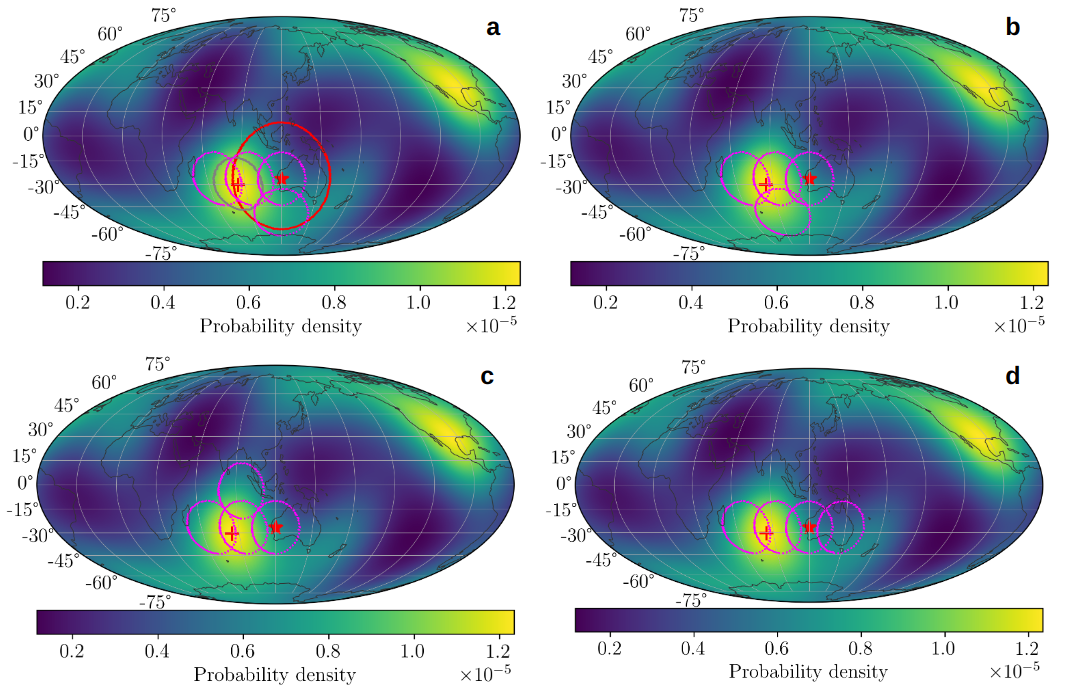}
\caption{
Similar to Figure~\ref{GW_map}, here we plot different observing strategies over the GW probability density map.
In Panel (a), the lines in different colors show the MWA FoV (down to 20\% of the maximum power) for different observing modes, with the red line corresponding to a single dipole per tile with maximum sensitivity at zenith, the gray line to the full array pointing to the most likely location of GW detections (red cross), and the magenta line to the four sub-arrays. The four pointings of the sub-array observing mode overlap at 50\% of the primary beam response, and one of them (the rightmost close to the equator) is towards the zenith. The red, gray, and magenta contours cover 12.4\%, 4.9\%, and 12.6\% of GW detections, respectively. Panels (b), (c), and (d) show different sub-array configurations in aid of determining the optimal pointings of the sub-array observing mode (see Section~\ref{sec:strategies}). 
}
\label{prob_map}
\end{figure*}

\begin{table*}
\centering
\caption{Pointings of the different observing mode beam configurations (see Section~\ref{sec:radio_criterion} and Figure~\ref{prob_map}) in the MWA frame in Alt/Az coordinates. Note that the full array and the single dipole per tile have a single beam, and the sub-arrays have four beams.}
\begin{tabular}{l|*{4}{c}}\hline
Mode
& Pointing\#1 & Pointing\#2 & Pointing\#3 & Pointing\#4\\
\hline
Full array &$(58.48^\circ, 261.21^\circ)$&&&\\
Single dipole per tile & $(90^\circ, 0^\circ)$ &  &&\\
Sub-arrays a &$(90^\circ, 0^\circ)$&$(66.85^\circ, 270^\circ)$&$(43.97^\circ, 270^\circ)$&$(66.85^\circ, 180^\circ)$\\
Sub-arrays b &$(90^\circ, 0^\circ)$&$(66.85^\circ, 270^\circ)$&$(43.97^\circ, 270^\circ)$&$(59.35^\circ, 219.88^\circ)$\\
Sub-arrays c &$(90^\circ, 0^\circ)$&$(66.85^\circ, 270^\circ)$&$(43.97^\circ, 270^\circ)$&$(56.09^\circ, 314.59^\circ)$\\
Sub-arrays d &$(90^\circ, 0^\circ)$&$(66.85^\circ, 270^\circ)$&$(43.97^\circ, 270^\circ)$&$(66.85^\circ, 90^\circ)$\\
\hline
\end{tabular}
\label{MWA_pointings}
\end{table*}

We plot the FoV (down to 20\% of the maximum power) of the three MWA observing modes in Panel (a) of Figure~\ref{prob_map}, with the red, gray, and magenta ellipses
corresponding to the single dipole per tile, full array, and an example four sub-arrays configuration, respectively. As shown in this figure, the pointing strategies for the three observing modes are different. While the single dipole per tile has maximum sensitivity at zenith, we chose to point the full array and the sub-arrays towards the O4 LVK 
highest sensitivity sky region above the Indian Ocean
(for our observing strategy see Section~\ref{sec:strategies}). 
The single dipole per tile, full array and sub-arrays can cover 12.4\%, 4.9\%, and 12.6\% of GW detections, respectively. As a comparison of sky coverage, we plot three more sub-array configurations, as shown in Panels (b), (c), and (d) of Figure~\ref{prob_map}, which will be used to determine the optimal pointings of the sub-array observing mode (see Section~\ref{sec:strategies}). The Alt/Az pointings of each of the six tested observing mode beam configurations are listed in Table~\ref{MWA_pointings}.

In order to infer the radio detectable fraction of GW events by the MWA, we convolved the probability coverage of the three observing modes with the fraction of GW and radio joint detections from our simulation (see Section~\ref{sec:GW_radio_joint}).
First, we took into account the sensitivity change over the MWA primary beam by creating a similar sensitivity map as shown for the LVK in Figure~\ref{prob_map} down to 20\% of the maximum power for the three different observing modes outlined in Section~\ref{sec:radio_criterion} (for the beam response of each MWA observing mode explored in this analysis see Figure~\ref{power_map} and Appendix~\ref{appendix:beam_response}).
Note that in the sub-array observing mode where the primary beams overlap 
we compared the responses from all beams and chose the best one for the overlapping regions
(see Panel c, d, e and f in Figure~\ref{power_map}). 
Next, for each position in the MWA primary beam, we applied a radio detection criterion with the corresponding fluence or flux density limit to obtain a radio detection fraction for each of the four models in Section~\ref{sec:models} based on our simulation results presented in Section~\ref{sec:GW_radio_joint}. Multiplying this fraction by the GW detection probability, i.e. the fraction of BNS mergers expected to be detected at that position by LVK during O4, and integrating over the MWA beam, we obtained the total number of GW and radio joint detections for the six observing mode beam configurations listed in Table~\ref{MWA_pointings}. 
The final detectable fraction of coherent radio emission from BNS mergers 
are summarised in Table~\ref{results}, which will be used to justify our observing strategies (see Section~\ref{sec:strategies}). We did a similar analysis for 200\,MHz and include the results in Table~\ref{results_200MHz} in Appendix~\ref{appendix:results_200MHz}.

As can be seen from Table~\ref{results}, the sub-array observing mode is expected to yield the most radio detections of BNS mergers 
due to a reasonable balance between sensitivity and sky coverage. While the full array has the best sensitivity, allowing a good fraction of the faint emission predicted by the magnetar collapse model to be detected, it cannot compete with the sub-arrays in regard to other emission models, especially the pulsar emission, which is so bright that sky coverage plays the dominant role. On the contrary, the single dipole has the largest sky coverage that can detect the most pulsar emission. However, its poor sensitivity is insufficient for detecting the emission from the magnetar collapse. In conclusion, sub-arrays are the best observing mode for searching for coherent radio emission from LVK O4 BNS mergers. 
Comparing the simulated results for the four sub-array configurations in Table~\ref{results}, we find that sub-array (b) has the highest probability for detecting signals predicted by the 
four emission models and is thus the optimal configuration to use.

\begin{table*}
\centering
\caption{Detectable fraction of the four model emissions at 120\,MHz (see Section~\ref{sec:models}) among all GW detections in O4 by the three MWA observing modes (see Section~\ref{sec:strategies}). The Sub-arrays a, b, c and d correspond to the four sub-array configurations displayed in Figure~\ref{prob_map}. \textbf{The bold row of Sub-arrays b is our preferred observing mode for searching for coherent radio emission from BNS mergers, as discussed in Section~\ref{sec:strategies}}.}
\begin{tabular}{l|*{4}{c}}\hline
\backslashbox{Mode}{Model}
& NS interaction & Jet-ISM interaction & Pulsar emission & Magnetar collapse\\
\hline
Full array &2.6\%&1.2\%&4.4\%&1.1\%\\
Single dipole per tile & 2.2\% & 0.2\% &11.2\%&0.1\%\\
Sub-arrays a &4.6\%&1.2\%&11.4\%&0.5\%\\
\setrow{\bfseries}Sub-arrays b &\setrow{\bfseries}4.6\%&\setrow{\bfseries}1.2\%&\setrow{\bfseries}11.8\%&\setrow{\bfseries}0.5\%\\
Sub-arrays c &4.2\%&1.1\%&11.2\%&0.5\%\\
Sub-arrays d &4.2\%&1.2\%&10.6\%&0.5\%\\
\hline
\end{tabular}
\label{results}
\end{table*}

\section{MWA observing strategies}\label{sec:strategies}

The MWA has a proven rapid-response system that can be on target within seconds of receiving a transient alert \citep{hancock19}, making it possible to catch any dispersion-delayed (FRB-like) radio signals emitted at the moment of a cataclysmic event by triggering observations with the high time resolution VCS. Combined with 
a wide FoV capable of covering a significant portion of the GW positional uncertainties, 
it is highly suitable for searching for coherent radio counterparts to BNS mergers. 
Here we discuss MWA observing strategies in targeting each of the four coherent emission models presented in Section~\ref{sec:models} based on our simulation results in Section~\ref{sec:simulation}. 
Our focus is on two observing systems 
available on the MWA, the rapid-response system and the buffering system. 
Our choice 
between these two observing systems 
is motivated by the expected arrival time of coherent radio emission according to the models presented in Section~\ref{sec:models} and the state of the telescope (whether it is observing or idle) at the time of an event.  
Note that the observing strategies discussed here are completely pre-programmed and automated via the Transient RApid-response using Coordinated Event Triggering (TRACE-T)\footnote{\url{https://github.com/ADACS-Australia/TraceT}} web application built under the Astronomy Data and Computing Services (ADACS) Merit Allocation Program (project IDs: GAnderson\_2022A, GAnderson\_2023A). 

The first two models presented in Sections~\ref{sec:model_NSB} and \ref{sec:model_ism}, which include the NS interaction and the jet-ISM interaction scenarios, predict prompt FRB-like emission to be emitted just prior to or during the merger. 
However, even if a BNS merger was detected at the nominal O4 and O5 horizon limit (190 and 300\,Mpc; \citealt{Abbott20}), the dispersion delay would be between $\lesssim14\text{--}40$\,s and $\lesssim22\text{--}60$\,s, respectively, at 120\,MHz depending on the Galactic and host dispersion measure contributions. An event that occurred at the same distance as GW170817 (40 Mpc) would have an even smaller delay of $\lesssim3\text{--}30$\,s \citep{James19}. 
These dispersion delays are potentially much shorter than the delay in the LVK pipeline produced Preliminary Alerts (delay of minutes), which will be the first automatic alert to contain positional information\footnote{\url{https://emfollow.docs.ligo.org/userguide/}}.
Even with a best-case rapid response time \citep[$<14$\,s;][]{hancock19}, it is unlikely the MWA will be on-target to detect the earliest predicted FRB-like signals if we rely solely on the Preliminary Alerts. 


In order to overcome this expected large latency, we plan to trigger MWA observations in a pre-specified observing configuration on any LVK alert of a new merger involving at least one NS component. 
On receiving an alert, the MWA will automatically divide into four sub-array pointings (i.e. the sub-array observing mode; see Section~\ref{sec:radio_criterion}), and 
shadow a large area of the sky that overlaps the highest sensitivity region of the LVK network over the Indian Ocean. 
Based on our simulation results in Section~\ref{sec:GW_map}, the sub-array (b) in Figure~\ref{prob_map} is the optimal observing mode for targeting the prompt radio emission predicted to be produced by NS interactions and jet-ISM interactions (see Table~\ref{results}), with the four beam pointing directions given in Table~\ref{MWA_pointings}.

In addition, on receiving an LVK alert, we can also trigger a ring buffer voltage dump \citep{Morrison23}, collecting 240\,s of negative latency data to catch any early merger emission.
As mentioned in Section~\ref{sec:radio_criterion}, we anticipate that at least some mergers with an NS component will be detected during their inspiral, which will generate an Early Warning Alert. Such alerts are likely to be transmitted at the time of the merger, which means that when combined with our proposed MWA sub-array pointing configuration, we will be on target in time to detect the very earliest FRB-like signals. 



Using MWA to target the other two emission models, i.e. the persistent pulsar emission (see Section~\ref{sec:model_pers}) and the magnetar collapse (see Section~\ref{sec:model_col}) is less time-critical as we expect any associated signals to occur between
$\sim1000\text{--}10000$\,s post-merger.
This means that 
the expected detectable fraction of radio signals for these two emission models 
will improve far beyond what we expect for our fixed beam and sub-array pointings listed in Table~\ref{results} (close to the predictions in Section~\ref{sec:GW_radio_joint}) as MWA repoints according to updated positional information in subsequent LVK alerts. 
Depending on the size and shape of the GW error distributions, we will either repoint the four sub-arrays to continue to cover large portions of the sky or drop to a single beam that utilises the full sensitivity of the array.
In order to capture the FRB-like radio signal from the magnetar collapse that could occur $\sim10000$\,s post-merger, we will continue recording with the VCS for up to 3 hours.

In summary, we propose a two-pronged triggering strategy for the MWA during the LVK O4 run in order to maximise our chances of a successful detection of coherent radio counterparts to GW events. If the MWA is idle then we will:

\begin{itemize}
    \item Shadow the LVK network's highest sensitivity region over the Indian Ocean, maximising our sky coverage by pointing MWA using the 4 beam sub-array
configuration (b) defined in Table~\ref{MWA_pointings} (see also Figure~\ref{prob_map}b) but operating 
in a no-capture or no-archive mode;
    \item On receiving an Early Warning Alert or a Preliminary Alert of a BNS or BH-NS merger, we will trigger the buffering mode, obtaining up to 240\,s of negative latency data on a significant portion of the sky and record for a further 15 minutes;  
    \item If the source is a BNS merger and we receive subsequent alerts (Preliminary Alerts and Initial Alerts) within 3 hours post-merger we will either
    \begin{itemize}
        \item continue observing with the 4 beam sub-array configuration (b) if no sky map is provided or if only 1 GW detector detects the event (resulting in a poor position), recording up to 1 hour post-merger or
        \item repoint the 4 sub-array beams individually to cover the best GW sky positions in the Southern Sky if the sky map is generated using two or more GW detectors, recording up to 3 hours post-merger. 
    \end{itemize}
    \item Continue to repoint the 4 sub-array beams individually on receiving subsequent alerts 
    (Preliminary Alerts and Initial Alerts) if improved positions and/or positional errors become available, recording up to 3 hours post-burst; and
    \item Cancel the observation in the case of a Retraction Alert or if the GW event is outside the MWA sky for up to 3 hours into the future.
\end{itemize}

\noindent When the MWA is in use:

\begin{itemize}
    \item On receiving an Early Warning Alert or Preliminary Alert, we will instead override the current
observations. If there is no or poor localisation,  we will use the 4 beam sub-array configuration (b)
(see Figure~\ref{prob_map}b and Table~\ref{MWA_pointings}), recording with the VCS for 15 minutes for a BH-NS merger or for up to one hour post-merger for a BNS merger;
\item If a sky map is available with the alert that was generated using two or more GW detectors,
we will repoint the array to cover the positional uncertainties in the Southern sky, either repointing the four sub-arrays individually or a single beam depending on the positional accuracy, recording up to 3 hours post-merger.
\item Continue to repoint according to the above strategy on receiving subsequent alerts with updated positions for up to 3 hours post-merger. 
\end{itemize}


The final data product collected by the VCS from each GW trigger will be raw voltages. In the case where the GW event is not well localised, we will perform an incoherent single pulse search for dispersed signals \citep{Xue17} using the {\sc presto} software package \citep{Ransom01} to target prompt radio signals predicted by the NS interaction, jet-ISM interaction and magnetar collapse models.
In the case where a GW event is well localised (positional error region of a few square degrees or with an identified electromagnetic counterpart), we will perform coherent beamforming \citep[potentially at several positions, e.g.][]{Tian22c} before conducting single pulse dedispersion searches. 
We can also perform offline correlation of the VCS data to create images over longer integrations to search for persistent pulsar emission or other predicted long-lived coherent radio emission \citep[e.g.][]{Starling20,Tian22b}.

During O4, there will be between $\sim14\,\text{--}\,85$ BNS mergers per year (depending on the assumed merger rate, which is extremely uncertain; \citealt{Abbott23}), out of which $\sim20\%$ ($\sim3\,\text{--}\,17$ BNS mergers) will have an Early Warning Alert (detected 10\,s prior to merger; \citealt{Magee21}). 
However, the BNS mergers with Early Warning Alerts may be outside the MWA field of view. 
Given the high sensitivity GW sky region constantly monitored by the MWA covers 12.6\% of GW BNS detections (see Section~\ref{sec:GW_map}), we expect up to 2 events with Early Warning Alerts 
to be within the MWA 4 beam sub-array configuration (b) field of view during O4.
Note that for the BNS mergers without Early Warning Alerts, we will still trigger on Preliminary Alerts and/or Initial Alerts, which contain positional information for pointing the MWA. Assuming 30\% sky coverage of the MWA (corresponding to the red contour in Figure~\ref{GW_map}), we would expect to trigger on $\sim3\,\text{--}\,20$ of these events per year.


In summary, we expect to successfully trigger on $\sim5\,\text{--}\,22$ BNS mergers per year during O4, of which 2 might have Early Warning Alerts. If we assume that all four coherent emission mechanisms operate in all BNS mergers (see Section~\ref{sec:models}), then we can predict how many BNS mergers will be detected by the MWA. For the 2 triggers with Early Warning Alerts, based on the detectable fraction of the model emission given in table~\ref{results}, we expect to detect the early emission models, i.e. the NS interaction and the jet-ISM interaction, from $\lesssim1$ BNS merger. For up to 22 triggers with Preliminary Alerts and/or Initial Alerts, we expect to detect the late-time emission models, i.e. the pulsar emission and the magnetar collapse, from $\sim2$ and $\lesssim1$ BNS mergers, respectively.

\section{Conclusions}

In this paper, we have investigated the prospects of detecting coherent radio counterparts of GW events using MWA VCS triggered observations. We have considered four coherent emission models applicable to BNS mergers, including the interaction of NS magnetic fields, jet-ISM interaction, persistent pulsar emission, and magnetar collapse, which were extensively studied in previous works \citep{Rowlinson19, Rowlinson20, Rowlinson21, Anderson20, Tian22a, Tian22b}. However, different from previous works, here we have taken into account the viewing angle dependence of these coherent emission models and found their detectability is largely dependent on this parameter.

In order to determine the radio detectable fraction of GW events, we have performed a population synthesis of binary mergers randomly distributed within the LVK O4 detector horizon. We have considered three observing modes for the MWA, the single dipole per tile with the largest FoV, the full array with the best sensitivity, and splitting MWA into four sub-arrays
for obtaining the best coverage of the LVK O4 high sensitivity region over the Indian Ocean (see Figure~\ref{prob_map}). As a result of this work, we come to the following main conclusions:

\begin{enumerate}
    \item Comparing the simulated radio detections by the three observing modes (see Table~\ref{results}), we have shown that sub-arrays are the best compromise between sky coverage and sensitivity as MWA will be on-sky to detect coherent radio emission from 12.6\% of the BNS merger population detected by LVK during O4.
    \item The 4 sub-array configuration (b) with pointings given in Table~\ref{MWA_pointings} provides the best coverage of all configurations tested. Assuming all BNS mergers detected during O4 emit coherent radio signals to those explored in Section~\ref{sec:models}, then we expect to detect 4.6\%, 1.2\%, 11.8\% and 0.5\% of coherent radio emission from the NS interaction, jet-ISM interaction, pulsar emission and magnetar collapse, respectively, when observing in this beam configuration.
    \item We predict MWA will successfully trigger on between $\sim5\,\text{--}\,22$ 
    BNS mergers per year during O4, 
    2 of which may have Early Warning Alerts and be in the 4 beam sub-array configuration (b) field of view, and the rest have Preliminary Alerts and/or Initial Alerts and be above the MWA horizon. For all these triggers, including both Early Warning Alerts and Preliminary and/or Initial Alerts, we expect to detect coherent radio emission predicted to be produced by the NS interaction, jet-ISM interaction, pulsar emission, and magnetar collapse from $\lesssim1$, $\lesssim1$, $\sim2$, and $\lesssim1$ BNS mergers, respectively.
\end{enumerate}


The MWA, with its rapid-response triggering system and buffering mode, is currently one of the most competitive radio telescopes for performing rapid follow-up observations of GWs in searching for coherent radio emission associated with BNS mergers. Based on the timescales of the various coherent emission models relative to the evolution of a BNS merger, we have proposed a triggering strategy to target each of them. We will keep the MWA pointed at the high sensitivity GW sky region and trigger the buffering mode to target the NS interaction and jet-ISM interaction models, and continue recording with the VCS for up to 3 hours to target the persistent pulsar emission and magnetar collapse models. With up to two successful early triggers during O4, we could potentially make the first detection of coherent radio emission from BNS mergers or place significant constraints on the models. 

\textbf{
Looking forward to the future, 
the MWA will soon undergo an upgrade to Phase III where all 256 tiles will be connected to the correlator, which will double the sensitivity to the millisecond timescale signals we are searching for. Based on the model predictions in Section~\ref{sec:simulation}, this additional sensitivity 
will improve our chances of detection with the MWA by a factor of $\sim1.5$. Furthermore, our experiment with the MWA demonstrates the importance of incorporating rapid response and sub-array observing capabilities into other low-frequency facilities such as the SKA-Low, which will have superior instantaneous sensitivity on short timescales.
In addition, the ability to rapidly trigger observations that tile large portions of the sky with sub-array beams is useful for many transient science cases beyond GW astrophysics, particularly in the multi-messenger field 
when considering neutrino events, cosmic rays, and very high-energy (TeV) gamma-ray transients.}

\begin{acknowledgements}
This scientific work uses data obtained from Inyarrimanha Ilgari Bundara / the Murchison Radio-astronomy Observatory, operated by CSIRO. We acknowledge the Wajarri Yamatji People as the Traditional Owners and native title holders of the Observatory site. Support for the operation of the MWA is provided by the Australian Government (NCRIS), under a contract to Curtin University administered by Astronomy Australia Limited.

KG acknowledges support through Australian Research Council Discovery Project DP200102243. FHP is supported by a Forrest Research Foundation Fellowship and acknowledges the support of the Australian Research Council (ARC) Centre of Excellence for Gravitational Wave Discovery under grant CE170100004.

This work was supported by resources provided by the Pawsey Supercomputing Centre with funding from the Australian Government and the Government of Western Australia.


The following software and packages were used to support this work:
{\sc Astropy} \citep{TheAstropyCollaboration2013,TheAstropyCollaboration2018}, 
{\sc numpy} \citep{vanderWalt_numpy_2011}, 
{\sc scipy} \citep{Jones_scipy_2001}, 
{\sc matplotlib} \citep{hunter07}.
This research has made use of NASA's Astrophysics Data System. 
\end{acknowledgements}



\bibliographystyle{pasa-mnras}
\bibliography{bib}

\appendix

\section{Radio emission from jet-ISM interaction for a narrow jet}
\label{appendix:GRB_jet_6deg}

\begin{figure}
\centering
\includegraphics[width=.5\textwidth]{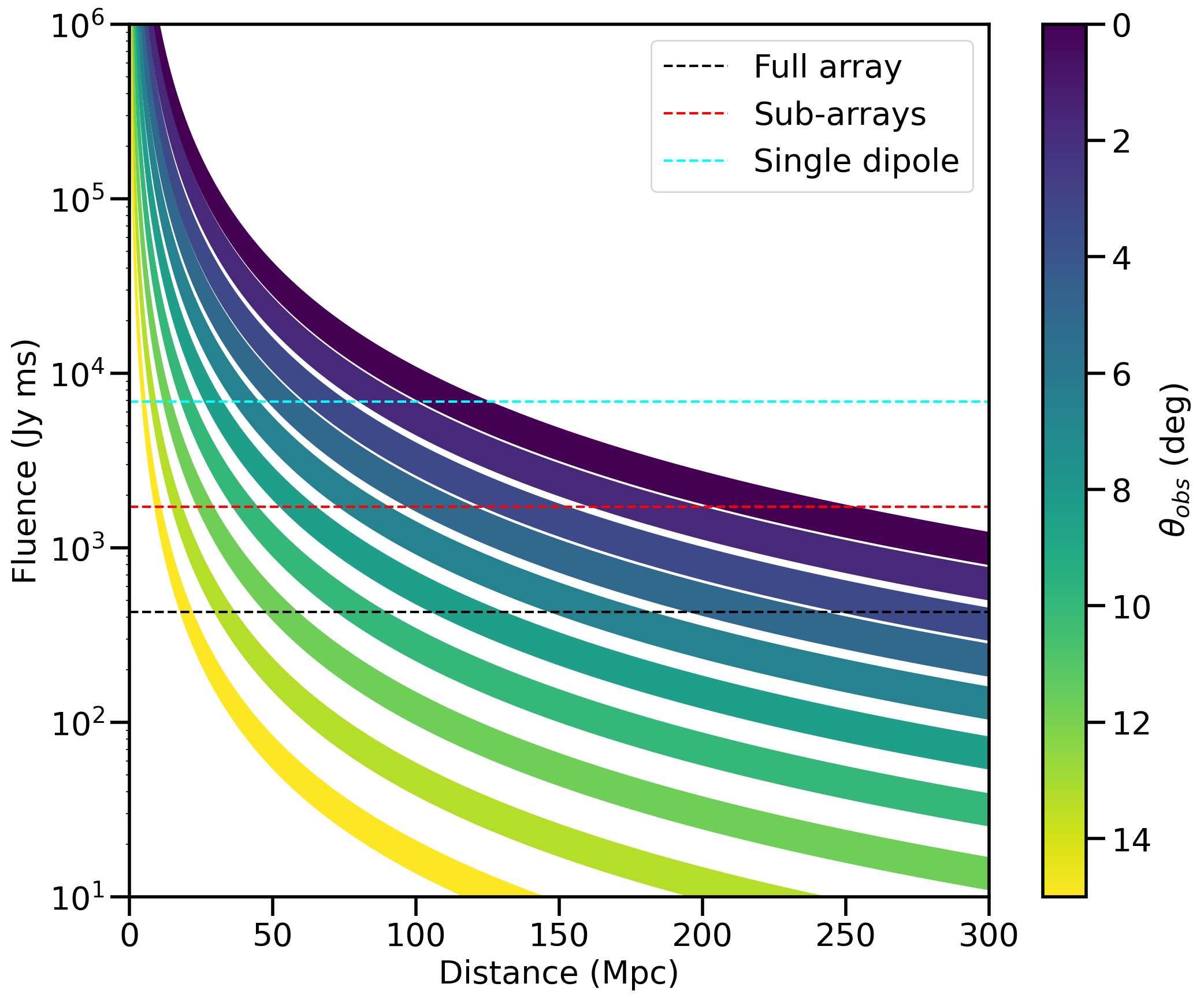}
\caption{Similar to Figure~\ref{GRB_jet_16deg}, here we plot the predicted radio emission from the jet-ISM interaction for a narrow jet with an opening angle of $\theta_0=\mathbf{6^\circ}$. The regions in different colors correspond to a range of viewing angles from $\mathbf{0^\circ}$ to $\mathbf{15^\circ}$.}
\label{GRB_jet_6deg}
\end{figure}

In Figure~\ref{GRB_jet_6deg}, we show the radio emission predicted to be produced by the relativistic jet-ISM interaction for a narrow jet for a range of viewing angles between $\mathbf{0^\circ}$ and $\mathbf{15^\circ}$. Here we adopt the same jet parameters as in Section~\ref{sec:model_ism} except the jet opening angle $\theta_0=\mathbf{6^\circ}$.

\section{Beam response for different observing modes}
\label{appendix:beam_response}

\begin{figure*}
\centering
\includegraphics[width=\textwidth]{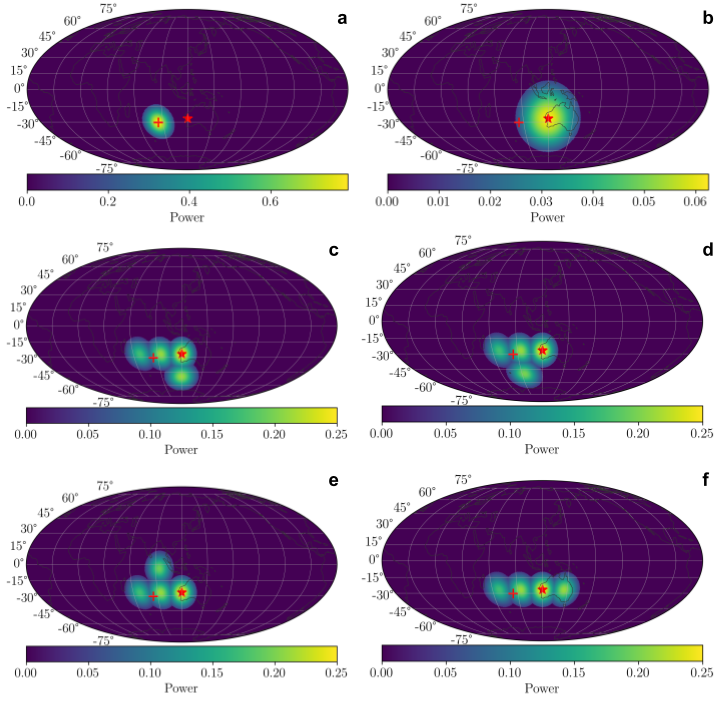}
\caption{Beam response at 120\,MHz for different observing modes displayed in Figure~\ref{prob_map}: (a) full array; (b) single dipole per tile; (c) sub-array a; (d) sub-array b; (e) sub-array c; (f) sub-array d. The power of the MWA full array at the zenith pointing is set to unity. We plot the response down to 20\% of each primary beam and ignore all sidelobes. For the overlapping region of primary beams in Panel c, d, e and f we compare the responses from all beams and plot the best one at each position. The red plus and star are the same as for Figure~\ref{GW_map}. Note that we have not plotted the primary beam side lobes responses, which would contribute a small amount of sensitivity to other parts of the sky but are less reliably calibrated.}
\label{power_map}
\end{figure*}

In Figure~\ref{power_map}, we show the beam response of the MWA for different beam observing configurations investigated in this paper (see Table~\ref{MWA_pointings}) with the aim of searching for coherent radio counterparts of GW events. These beam response maps were used to calculate the varying sensitivity of MWA across the LVK O4 GW sensitivity map (see Figure~\ref{prob_map}) and predict the rate of GW-radio joint detections (see Section~\ref{sec:GW_map}).

\section{Radio detections of GW events at 200MHz}
\label{appendix:results_200MHz}

In Table~\ref{results_200MHz}, we provide the rate of detecting radio counterparts to GW events predicted by the four coherent emission models (see Section~\ref{sec:models}) for the three MWA observing modes (see Section~\ref{sec:strategies}) at 200\,MHz. Compared to the results at 120\,MHz (see Table~\ref{results}), the detection rate here is much lower for two reasons: the smaller FoV of MWA at 200\,MHz (see Table~\ref{mode_summary}); and the predicted emission being fainter at 200\,MHz. However, at both 120\,MHz and 200\,MHz the sub-array observing mode has the highest radio detection rate and thus is the optimal observing strategy (see Section~\ref{sec:strategies}).

\begin{table*}
\centering
\caption{Detectable fraction of the four model emissions at 200\,MHz (see Section~\ref{sec:models}) among all BNS detections in O4 by the three MWA beam observing configurations listed in Table~\ref{MWA_pointings} (see also Section~\ref{sec:strategies}). The Sub-arrays a, b, c and d correspond to the four sub-array configurations displayed in Figure~\ref{prob_map}.}
\begin{tabular}{l|*{4}{c}}\hline
\backslashbox{Mode}{Model}
& NS interaction & Jet-ISM interaction & Pulsar emission & Magnetar collapse\\
\hline
Full array &0.77\%&0.2\%&1.32\%&0.03\%\\
Single dipole per tile & 1.39\% & 0.04\% &7.56\%&$<0.01\%$\\
Sub-arrays a &1.32\%&0.12\%&3.22\%&0.01\%\\
Sub-arrays b &1.39\%&0.14\%&3.37\%&0.01\%\\
Sub-arrays c &1.3\%&0.13\%&3.23\%&0.01\%\\
Sub-arrays d &1.19\%&0.12\%&3.02\%&0.01\%\\
\hline
\end{tabular}
\label{results_200MHz}
\end{table*}

\end{document}